\documentclass[%
 preprint,
]{revtex4-2}

\usepackage{graphicx}
\usepackage{multirow}
\usepackage{subcaption}
\usepackage{dcolumn}
\usepackage{bm}
\usepackage{xcolor} 

\usepackage{amsmath}
\usepackage{amssymb}
\usepackage{booktabs}
\usepackage{natbib}
\usepackage{xcolor} 

\begin{document}


\title{Stabilizing Rayleigh–Bénard convection with reinforcement learning trained on a reduced-order model
}

\author{Qiwei Chen and C. Ricardo Constante-Amores}
\affiliation{Department of Mechanical Science and Engineering, University of Illinois, Urbana Champaign, USA}

\begin{abstract}

Rayleigh–Bénard convection (RBC) is a canonical system for buoyancy-driven turbulence and heat transport, central to geophysical and industrial flows. Developing efficient control strategies remains challenging at high Rayleigh numbers, where fully resolved simulations are computationally expensive. We use a control framework that couples data-driven manifold dynamics (DManD) with reinforcement learning (RL)  to suppress convective heat transfer. We find a coordinate transformation to a low-dimensional system  using POD and autoencoders, and then  learn an evolution equation for this low-dimensional state using  neural ODEs. The reduced model reproduces key system features  while enabling rapid policy training.
Policies trained in the DManD environment and deployed in DNS achieve a 16–23\% reduction in the Nusselt number for both single- and dual-boundary actuation. Physically, the learned strategy modulates near-wall heat flux to stabilize and thicken the thermal boundary layer, weaken plume ejection, and damp the wall-driven instabilities that seed convective bursts. Crucially, the controller drives the flow toward a quasi-steady state characterized by suppressed temporal fluctuations and spatially steady heat-flux patterns.
This work establishes DManD–RL as a physically interpretable, scalable approach for turbulence control in high-dimensional flows. 
\end{abstract}

\maketitle

\section{Introduction}
Rayleigh–Bénard convection (RBC) is a classical system in fluid dynamics, which is characterized by buoyancy-driven flow patterns arising from  a temperature gradient~\citep{ahlers2009heat}. While RBC has long served as a canonical model for studying nonlinear dynamics and pattern formation, its control has become an essential research problem with both theoretical and practical importance~\citep{bodenschatz2000recent, lohse2010small, chillà2012new}. Effective control of convective flows is crucial in engineering applications where heat transport plays a dominant role, such as energy efficiency in buildings and temperature regulation in  chemical  or industrial processes~\citep{rahimi2019comprehensive, guo2025optimization}. Beyond its technological relevance, RBC also provides a simplified framework for understanding buoyancy-driven phenomena in nature, including 
atmospheric and oceanic circulations, mantle convection, and stellar dynamics, where buoyancy-driven instabilities govern large-scale transport~\citep{kadanoff2001turbulent, markson1975atmospheric, nordlund2009solar}. Thus, studying RBC control strategies is essential not only for improving practical systems, but also for deepening our understanding of fundamental fluid dynamical processes.

Traditional approaches for RBC control have primarily focused on model-based or experimental feedback approaches. ~\citet{or2001robust} employed a linear–quadratic–Gaussian feedback controller to stabilize the unstable (no-motion) state, while ~\citet{howle1997active} implemented active thermal actuation in laboratory experiments. Although these methods demonstrated partial flow stabilization, their effectiveness was limited to moderate conditions. More recent work has introduced reinforcement learning (RL) as an alternative,
which can autonomously discover nonlinear control policies and achieve stronger stabilization of chaotic states ~\citep{beintema2020controlling, vignon2023effective, jeon2025inductive}. However, RL training typically requires continuous interaction with fully resolved direct numerical simulations (DNS) of the governing equations. Such a tight coupling between the learning algorithm and DNS makes training prohibitively costly. In particular, at high Rayleigh numbers, i.e. $Ra = 10^6$, the turbulent dynamics require fine spatial resolution and small time steps to resolve all the relevant flow scales, rendering direct interaction between control algorithms and fully resolved simulations computationally expensive.

In recent years, data-driven manifold dynamics (DManD) has emerged as a effective approach to obtain reduced order models for complex dynamics \citep{Linot_Graham_2023}. The framework identifies a set of manifold coordinates that capture the essential dynamics of the system with far fewer degrees of freedom than the full state and then determines the governing equations for their evolution in the reduced space.
This approach has been successfully applied to a variety of systems, including the Kuramoto–Sivashinsky equation~\citep{linot2020deep}, two-dimensional Kolmogorov flow~\citep{constante2024data,CRCA_2024}, turbulent pipe flow~\citep{pipe_jfm}, and Rayleigh–Bénard convection~\citep{chen2025dynamics}. In each case, it  has accurately  reproduced the  essential nonlinear dynamics with high fidelity while achieving substantial dimension reduction. Because the temporal evolution in DManD occurs on a reduced space with far fewer degrees of freedom, it operates orders of magnitude faster than DNS. Consequently, combining DManD with RL
provides a promising route to accelerate the training while retaining physical fidelity. \citet{zeng2022data} first demonstrated this DManD-RL framework  for controlling spatiotemporal chaotic dynamics in the 1D Kuramoto–Sivashinksy equation, where the RL agent learned a policy that drove the system toward   a laminar state. \citet{linot2023turbulence} subsequently extended  this framework to 3D turbulent Couette flow, demonstrating its efficiency in more complex and high-dimensional scenarios where the RL learned a policy to drag the system toward a high dissipation state prior to relaminarization. These results suggest that the DManD-RL framework is a versatile and scalable tool for addressing flow control problems across a broad range of dynamical systems.

In this work, we stabilize Rayleigh–Bénard convection at $Ra=10^6$ by applying RL-based control strategies learned within the DManD framework to  fully resolved direct numerical simulations.
A low-dimensional DManD model of RBC dynamics is first trained, after which  RL is used to design control strategies on this ROM, enabling efficient policy  training while preserving high fidelity to the original dynamics. The learned policies are then deployed  in the full DNS environment, showing robust performance and effective manipulation of convective heat transfer. The remainder of this paper is organized as follows. Section~\ref{sec:formulation} introduces the details of the DManD-RL framework; Section~\ref{sec:results} presents the results for RBC control; and Section~\ref{sec:conclusion} summarizes our findings and discusses future directions.

\section{Formulation}
\label{sec:formulation}

\subsection{Direct numerical simulation of RBC}
\label{sec:DNS}
To generate data set for model learning and control evaluation, we simulate two-dimensional (2D) Rayleigh-Bénard convection with dimensionless governing equations:

\begin{equation}
\begin{aligned}
\nabla \cdot \boldsymbol{u} &= 0, \\
\frac{\partial \boldsymbol{u}}{\partial t} + (\boldsymbol{u} \cdot \nabla)\boldsymbol{u} &= -\nabla p + \sqrt{\frac{\mathrm{Pr}}{\mathrm{Ra}}}\,\nabla^2 \boldsymbol{u} + T\,\hat{\boldsymbol{z}}, \\
\frac{\partial T}{\partial t} + (\boldsymbol{u} \cdot \nabla) T &= \frac{1}{\sqrt{\mathrm{Ra}\,\mathrm{Pr}}}\,\nabla^2 T.
\end{aligned}
\end{equation}
where $\boldsymbol{u} = (u_x, u_y)$ is the velocity field, $T$ is the temperature, and $p$ is the pressure. The unit vector $\hat{\boldsymbol{z}}$ denotes the vertical direction. The Rayleigh number $\mathrm{Ra} = 10^6$ and  Prandtl number $\mathrm{Pr} = 1$. We choose this relatively high Rayleigh number because it lies well within the turbulent convection regime~\citep{ahlers2009heat, chillà2012new}, where strong plume emission~\citep{de2018dynamics}, coherent roll interactions~\citep{schneide2019lagrangian}, and multiscale behaviors~\citep{pandey2018turbulent} characterize the flow dynamics. In such a regime, even small perturbations can cause significant nonlinear transitions in local convection cells, making active control more challenging. 

The computational domain is $[0, \pi] \times [0, 1]$, discretized with $96 \times 64$ spectral grid points using a Fourier–Chebyshev basis. This resolution is adopted in other RBC modeling and controlling works~\citep{10651496,markmann2025controlrayleighbenardconvectioneffectiveness}. The thermal boundary conditions are Dirichlet: $T = 1 + \epsilon$ at the bottom boundary ($y = 0$) and $T = \epsilon$ at the top boundary ($y = 1$), where $\epsilon \in [0, 0.75]$ denotes the control action. The velocity field satisfies no-slip boundary conditions, i.e., $\boldsymbol{u} = 0$ at $y = 0$ and $y = 1$. 
Periodic boundary conditions are applied in the streamwise direction.
The simulations are implemented using the Dedalus framework ~\citep{2020PhRvR...2b3068B}, employing a dealiasing factor of $3/2$ and second-order RK222 scheme. The maximum timestep is $\Delta t = 0.125$, and flow snapshots are saved every $\Delta t = 0.5$. Each simulation is run until a final time of $t = 225$. After discarding the initial warm-up period ($t \leq 25$), we retain 400 snapshots per trajectory for training and analysis.

The instantaneous Nusselt number, Nu, is computed during the simulations as a diagnostic quantity and control target. This is defined as
\begin{equation}
\mathrm{Nu}(t) = \sqrt{\mathrm{Ra} \cdot \mathrm{Pr}} \, \langle u_z T \rangle - \left\langle \frac{\partial T}{\partial z} \right\rangle,
\end{equation}
where $u_z$ is the vertical velocity, $T$ is the temperature field, and $\langle \cdot \rangle$ denotes a spatial average over the domain. The $Nu$ number quantifies the efficiency of convective heat transport. Lower $Nu$ numbers correspond to reduced convective motion  and a state approaching purely diffusive heat transfer.

\begin{figure}[t]
  \centering
\includegraphics[width=\textwidth]{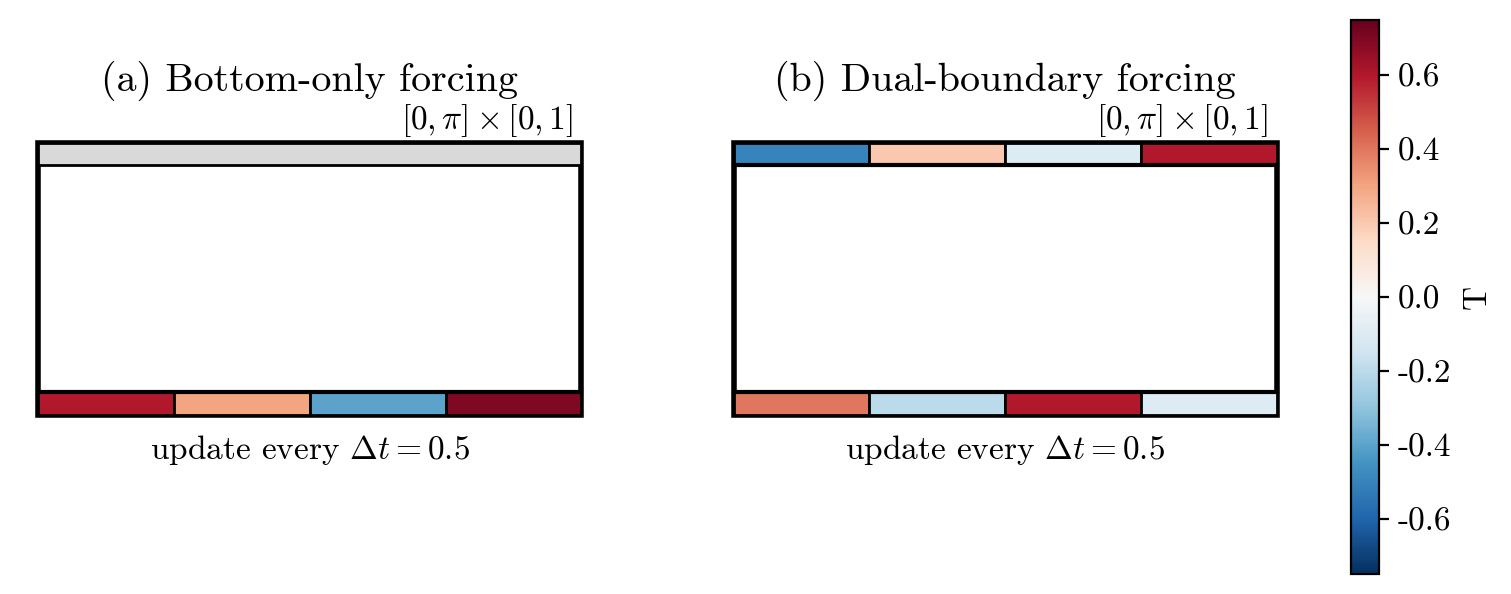}
  \caption{Framework of boundary-forcing configurations used to generate the dataset.}
  \label{fig:boundary_framework}
\end{figure}

Two simulation configurations are used to generate training data under random boundary forcing, corresponding to different control schemes. The first configuration applies perturbations only at the bottom boundary. The lower wall is divided into four equal segments, each updated every $\Delta t = 0.5$. During the warm-up stage, the boundary temperature is fixed at 1.0. Afterward, each segment is randomly sampled from the uniform range $[1 - \epsilon, 1 + \epsilon]$ with $\epsilon = 0.75$, introducing localized heating variations.
The second configuration extends this setup by also perturbing the top boundary. Both the bottom and top boundaries are divided into four segments; after the initial fixed period, the bottom segments are sampled from $[1 - \epsilon, 1 + \epsilon]$, while the top segments are drawn from $[-\epsilon, \epsilon]$. This dual-boundary forcing introduces more diverse thermal patterns and richer convective dynamics~\citep{bassani2022rayleigh}. A schematic of the single- and dual-boundary forcing configurations is shown in Figure~\ref{fig:boundary_framework}. For each case, we simulate 200 initial conditions, resulting in a total of 80,000 snapshots per control scheme for model training.

\subsection{Learning manifold coordinates}

\subsubsection{Linear reduction via POD}
\label{sec:pod}
We apply snapshot POD~\citep{annurev:/content/journals/10.1146/annurev.fl.25.010193.002543} to the flattened flow fields $(u_x, u_y, T)$ to extract orthonormal spatial modes that best capture the dataset variance. Let $\mathbf{u}(t) \in \mathbb{R}^D$ denote a snapshot vector at time $t$, where $D = 3 \times N_x \times N_y = 3\times96\times64$. 
The dataset is organized as a matrix $\mathbf{U} = [\mathbf{u}(t_1), \mathbf{u}(t_2), \dots, \mathbf{u}(t_N)]^\top \in \mathbb{R}^{N \times D}$, where $N$ is the number of snapshots.

To do POD, we define the state variable as $\boldsymbol{q}(\boldsymbol{x}) = [\boldsymbol{u}(\boldsymbol{x}), T(\boldsymbol{x})]$, where $\boldsymbol{u}$ is the velocity field and $T$ is the temperature field. The aim of POD is to find a function $\boldsymbol{\Phi}$ that maximizes
\begin{equation}
    \frac{\left \langle \left |  (\boldsymbol{q}',\boldsymbol{\Phi})_E\right |^2  \right \rangle }{|| \boldsymbol{\Phi}  ||^2_E  },
\end{equation}
where $\boldsymbol{q}'(\boldsymbol{x})= \boldsymbol{q}(\boldsymbol{x}) - \bar{\boldsymbol{q}}(\boldsymbol{x})$ stands for the fluctuating component of the state variable, and $\bar{\boldsymbol{q}}$ is the mean over time. $\left \langle \cdot \right \rangle$ denotes the ensemble average, and the inner product is defined to be
\begin{equation}\label{eigenvalue1}
(\boldsymbol{q}_1,\boldsymbol{q}_2)_E = \int_\Omega \boldsymbol{q}_1 \cdot \boldsymbol{q}_2 \, d\boldsymbol{x},
\end{equation}
with the corresponding energy norm $|| \boldsymbol{q} ||_E = (\boldsymbol{q}, \boldsymbol{q})_E$.

The POD basis functions $\boldsymbol{\Phi}^{(n)}(\boldsymbol{x})$ are obtained by solving the eigenvalue problem associated with the space-time covariance of the fluctuating state field $\boldsymbol{q}' = \boldsymbol{q} - \bar{\boldsymbol{q}}$:
\begin{equation}
\int_{\Omega} \left\langle \boldsymbol{q}'(\boldsymbol{x},t) \otimes \boldsymbol{q}'(\boldsymbol{x}',t) \right\rangle \boldsymbol{\Phi}^{(n)}(\boldsymbol{x}') \, d\boldsymbol{x}' = \lambda^{(n)} \boldsymbol{\Phi}^{(n)}(\boldsymbol{x}),
\end{equation}
where $\lambda^{(n)}$ are the POD eigenvalues corresponding to the energy captured by each mode.

To obtain these modes, we first organize the fluctuating state field as 
\begin{equation}
    \boldsymbol{X} = [\boldsymbol{q}'_0, \boldsymbol{q}'_1, ...,\boldsymbol{q}'_M ] \in \mathbb{R}^{N \times M}
\end{equation}
Then the empirical covariance operator is approximated as
\begin{equation}
    \mathbf{C} = \frac{1}{M - 1} X X^\top \in \mathbb{R}^{N \times N}.
\end{equation}
Direct eigendecomposition of $\mathbf{C}$ is computationally expensive for large $N$, so we apply the method of snapshots by instead solving the smaller eigenvalue problem
\begin{equation}
    \tilde{\mathbf{C}} \mathbf{v}^{(n)} = \lambda^{(n)} \mathbf{v}^{(n)}, \quad \text{where } \tilde{\mathbf{C}} = \frac{1}{M - 1} X^\top X \in \mathbb{R}^{M \times M}.
\end{equation}
The spatial POD modes are then reconstructed by projecting back to the full space:
\begin{equation}
    \boldsymbol{\Phi}^{(n)} = \frac{X \mathbf{v}^{(n)}}{\left\| X \mathbf{v}^{(n)} \right\|},
\end{equation}
where the denominator ensures $L^2$ normalization. This yields a set of orthonormal modes $\{ \boldsymbol{\Phi}^{(n)} \}$ capturing the most energetic structures in the dataset.

The POD coefficient $\mathbf{a}_i(t)$ is obtained by projecting the fluctuating field
onto the spatial mode $\boldsymbol{\Phi}^{(n)}$:
\begin{equation}
    \mathbf{a}_i(t) = \int_{\Omega} \boldsymbol{q}'(\boldsymbol{x},t) 
    \cdot \boldsymbol{\Phi}^{(i)}(\boldsymbol{x}) \, d\boldsymbol{x}.
\end{equation}

\subsubsection{Nonlinear reduction via autoencoders}
\label{sec:autoencoder}
To further reduce  the dimension of the system, we apply a fully connected autoencoder to the truncated set of POD coefficients. The encoder compresses the POD coefficients into a lower-dimensional latent vector $\boldsymbol{h} \in \mathbb{R}^{d_h} $, while the decoder reconstructs them back into the original space. The loss function is defined as the mean-squared reconstruction error between the input and output POD coefficients
\begin{equation}
    \mathcal{L}(\theta) = \frac{1}{N} \sum_{i=1}^N \left\| \mathbf{a}_i(t) - \mathcal{D}(\mathcal{E}(\mathbf{a}_i(t))) \right\|^2_2,
\end{equation}
where $\mathcal{E}$ and $\mathcal{D}$ denote the encoder and decoder networks with parameters \(\theta\). The training data are normalized by subtracting the temporal mean and dividing by the maximum standard deviation to improve numerical stability. 

The network architecture and training configuration are summarized in Table~\ref{tab:ae-architecture}. The model is trained for  500 epochs using the AdamW~\citep{kingma2014adam} optimizer with an initial learning rate of $1 \times 10^{-3}$ and a weight decay of $1 \times 10^{-6}$. A StepLR scheduler reduces the learning rate by a factor of 0.5 every quarter of the total training iterations. Both the encoder and decoder use GELU activation in all hidden layers, and a final linear output layer.

\begin{table}
\centering
\caption{Architecture of the autoencoder used in DManD.}
\label{tab:ae-architecture}
\begin{tabular}{cccc}
\toprule
Function & Shape & Activation & Learning Rate \\
\midrule
Encoder $\mathcal{E}$ & $a_{\text{dim}}$ / 800 / 800 / 800 / $h_{\text{dim}}$ & GELU / GELU / GELU / Linear & $1 \times 10^{-3}$ (StepLR) \\
Decoder $\mathcal{D}$ & $h_{\text{dim}}$ / 800 / 800 / 800 / $a_{\text{dim}}$ & GELU / GELU / GELU / Linear & $1 \times 10^{-3}$ (StepLR) \\
\bottomrule
\end{tabular}
\end{table}

\subsection{Learning NODE in actuated dynamics}
To model the temporal evolution of the low dimensional space   $\mathbf{h}(t)$, we employ a neural ordinary differential equation (NODE) model~\citep{chen2018neural}. The NODE learns the mapping
\[
\frac{d\mathbf{h}}{dt} = f(\mathbf{h}, \mathbf{a}_{\mathrm{ctrl}}),
\]
where $\mathbf{a}_{\mathrm{ctrl}}$ is the external control input. Importantly, the NODE only predicts the evolution of $\mathbf{h}$; the control $\mathbf{a}_{\mathrm{ctrl}}$ is treated as a non-dynamic external input and is not evolved during training or rollout.

The neural network $f$ is implemented as a fully connected multilayer perceptron (MLP) with GELU activations and four hidden layers, as summarized in Table~\ref{tab:node-arch}. The network takes the concatenated $(\mathbf{h}, \mathbf{a}_{\mathrm{ctrl}})$ as input and returns the latent derivative $\dot{\mathbf{h}}$, with a residual output of zero for the control variable.

\begin{table}
\centering
\caption{Architecture of the NODE 
}
\label{tab:node-arch}
\begin{tabular}{cccc}
\toprule
Input & Hidden Layers & Output & Activation \\
\midrule
$h \in \mathbb{R}^{h_{\text{dim}}},\ a \in \mathbb{R}^{\text{control}_{\text{dim}}}$ & 600 / 600 / 600 / 600 & $\dot{h} \in \mathbb{R}^{h_{\text{dim}}}$ & GELU / GELU / GELU / Linear \\
\bottomrule
\end{tabular}
\end{table}

The model is trained using the Adam optimizer with a learning rate of $2.5 \times 10^{-5}$ and a cosine annealing learning rate scheduler. Training is performed on batches of short temporal segments ($\text{batch\_time}=2$) sampled from multiple trajectories, using a mean squared error loss on predicted latent sequences. The ODE integration is solved using the \texttt{torchdiffeq} package with the Dormand–Prince method (``dopri5'') and tolerances of $10^{-6}$ (rtol) and $10^{-8}$ (atol).

Once trained, the NODE can be used to evolve  latent trajectories $\mathbf{h}(t)$ under prescribed control inputs $\mathbf{a}_{\mathrm{ctrl}}(t)$, without the need to query the full DNS. This formulation enables highly efficient surrogate modeling and facilitates reinforcement learning directly within the low dimensional representation.

\subsection{Learning control strategy in the low dimensional model}

After the DManD model is trained and validated, reinforcement learning is performed entirely within this low-dimensional model  rather than the full DNS. 
This approach accelerates control learning, as the evolution of the full high-dimensional flow field is replaced by the time evolution of a low-dimensional dynamical system that captures the essential large-scale flow structures.

At each control interval $\Delta t_{\mathrm{ctrl}} = 0.5$, the RL agent receives the current latent state $\mathbf{h}_t \in \mathbb{R}^{d_h}$ (obtained by encoding the POD coefficients through the autoencoder), and outputs a control vector
\[
\mathbf{a}_{\mathrm{ctrl}} = \pi_\theta(\mathbf{h}_t) \in \mathbb{R}^{n_{\mathrm{seg}}},
\]
where $n_{\mathrm{seg}} = 4$ for single-boundary and $8$ for dual-boundary control.  
These actions are passed to the DManD model $\frac{d\mathbf{h}}{dt} = f(\mathbf{h}, \mathbf{a}_{\mathrm{ctrl}})$, yielding the next state $\mathbf{h}_{t+1}$.  
The agent therefore explores and learns directly in the latent space governed by $f$, avoiding the computational cost of DNS integration. Importantly, while $\Delta t_{\mathrm{ctrl}}$ is set to 0.5 in this numerical study, it can be adjusted to a smaller value in practical applications to accommodate the finite response time and the constrain of speed of system control.

The reward function is formulated  to minimize heat transfer, measured by the instantaneous Nusselt number, while penalizing excessive control effort
\begin{equation}
    r_t = -\,\text{Nu}_t - \lambda \|\mathbf{a}_{\mathrm{ctrl}} - \mathbf{a}_{\text{base}}\|^2,
\end{equation}
where $\lambda$ is a tunable penalty coefficient and $\mathbf{a}_{\text{base}}$ denotes the no-control action. This penalty term accounts for a key practical constraint regarding the actuation magnitude, preventing the controller from demanding unphysically high energy inputs. 
The agent is trained using the Twin Delayed Deep Deterministic Policy Gradient (TD3) algorithm~\citep{fujimoto2018addressingfunctionapproximationerror}, 
which employs two critic networks $Q_{\phi_1}$ and $Q_{\phi_2}$, along with  one actor $\pi_\theta$ to ensure training stability.  
The $Q_{\phi_1}$ and $Q_{\phi_2}$ networks are optimized via the temporal-difference objective
\begin{equation}
\mathcal{L}_{\text{critic}}
= \frac{1}{N}\sum_{t=1}^{N}
\big( Q_{\phi_i}(\mathbf{h}_t, \mathbf{a}_{\mathrm{ctrl}})
- y_t \big)^2,
\qquad
y_t = r_t + \gamma \min_{i=1,2} Q_{\phi_i}\big(\mathbf{h}_{t+1}, \pi_\theta(\mathbf{h}_{t+1})\big),
\end{equation}
where the summation is taken over training samples when computing the loss. We employ temporally correlated Gaussian action noise for exploration and use the AdamW optimizer with a cosine learning rate schedule starting at $2.5\times10^{-4}$.  
The actor is implemented as an eight hidden layers fully connected network with ReLU activations, while each critic network follows the standard two-layer TD3 architecture.  
Table~\ref{tab:td3-architecture} summarizes the model configuration and training setup.

Once the control policy $\pi_\theta$ converges, it can be directly deployed in the full DNS as described in Section~\ref{sec:apply_policy_dns}. This latent-space formulation enables control learning to proceed several orders of magnitude faster than DNS integration, while retaining accurate representations of the flow's large-scale structures and feedback dynamics.

\begin{table}
\centering
\caption{TD3 architecture and hyperparameters used in the  DManD control training.}
\begin{tabular}{ccc}
\toprule
Component & Layer dimensions & Activation \\
\midrule
Actor (policy network) & 256 / $(256)^6$ / 256 & ReLU (each) \\
Critic networks ($Q_{\phi_1}, Q_{\phi_2}$) & 256 / 256 & ReLU (each) \\
Optimizer & AdamW & Cosine LR schedule ($2.5\times10^{-4}$ start) \\
Action noise & $\mathcal{N}(0,\,0.2^2)$ & temporally correlated \\
\bottomrule
\end{tabular}
\label{tab:td3-architecture}
\end{table}

\vspace{0.3em}

\subsection{Apply control strategy in the full DNS}
\label{sec:apply_policy_dns}

Next, the control policy trained within the DManD framework is implemented in closed-loop with the full DNS to assess its performance on the original high-dimensional flow.
At each control step ($\Delta t_{\mathrm{ctrl}} = 0.5$) the instantaneous DNS state
$\mathbf{q}_{n} \in \mathbb{R}^{N}$ is mapped to a  latent representation using the previously trained POD–autoencoder framework
\begin{equation} \mathbf{h}_n = \mathcal{E}(\int_{\Omega} (\boldsymbol{q}_n(\boldsymbol{x}) - \bar{\boldsymbol{q}}(\boldsymbol{x})) \cdot \boldsymbol{\Phi}^{(i)}(\boldsymbol{x}) \, d\boldsymbol{x}), \label{eq:get_h} 
\end{equation}
where $\bar{\boldsymbol{q}}(\boldsymbol{x})$ denotes the mean field from   the training set used in the POD decomposition.
The resulting latent vector $\mathbf{h}_n$ is normalized using the training statistics:
$\widehat{\mathbf{h}}_n = (\mathbf{h}_n - \bm{\mu}_h)/\bm{\sigma}_h$.
The trained actor network $\pi_\theta$ 
then produces the control action
\begin{equation} \mathbf{a}_{\mathrm{ctrl}}(t)\;=\;\pi_\theta\!\big(\widehat{\mathbf{h}_n}\big)\in\mathbb{R}^{n_{\mathrm{seg}}}, 
\end{equation} 
with $n_{\mathrm{seg}}=4$ segments ($n_{\mathrm{seg}}=8$ for dual-boundary control).
The components of $\mathbf{a}_{\mathrm{ctrl}}$ are assigned to the corresponding boundary segments over $x \in [0,\pi]$
\begin{equation} b(x,0,t)=a_{\mathrm{ctrl},i}(t). \end{equation}
The DNS is then advanced forward under these boundary conditions until the next control update at $t_{n+1} = t_n + \Delta t_{\mathrm{ctrl}}$, producing the next system state $\mathbf{q}_{n+1}$. This process is repeated iteratively following Eq.~\eqref{eq:get_h}, thereby establishing a fully coupled DNS–RL feedback loop. No online learning occurs, ensuring that observed flow modification arise purely from the trained control law interacting with the DNS dynamics.

\section{Results \label{sec:results}}

\subsection{Dimension reduction}

With this dataset, we begin by performing POD on the flow field snapshots as described in section \ref{sec:pod} to obtain an energy-ranked orthogonal basis for the system. For POD, a total of 8,000 flow snapshots is extracted by sampling every five time units.
Among these, 5,000 are used for extracting POD modes, and the remaining 3,000  for testing. The POD modes are ordered in descending order of their eigenvalues $\lambda_i$, which represent the kinetic and thermal energy content captured by each mode. The cumulative energy fraction retained by the leading $r$ modes is defined as
$\sum_{i=1}^r \lambda_i \Big/ \sum_{i=1}^{D} \lambda_i \geq \tau$,
where $\tau$ denotes the target energy threshold. In this work, we set $\tau = 99.95\%$, which yields $r=551$ modes for the single-boundary control case and $r=618$ modes for the dual-boundary control case. This truncation ensures that nearly all energetic and flow structures are preserved while maintaining computational tractability for for the subsequent nonlinear reduction. Furthermore, as discussed in \citet{Holmes_Lumley_Berkooz_1996} and \citet{epps2010error}, noise typically manifests as high-frequency, low-energy fluctuations that are preferentially captured by higher-order modes. 
As a result, projection to POD removes these contributions, and the resulting low-dimensional representation acts as an effective noise-rejection filter, enhancing the robustness of the reduced-order model.

\begin{figure}[htbp]
  \centering
   \includegraphics[width=\textwidth]{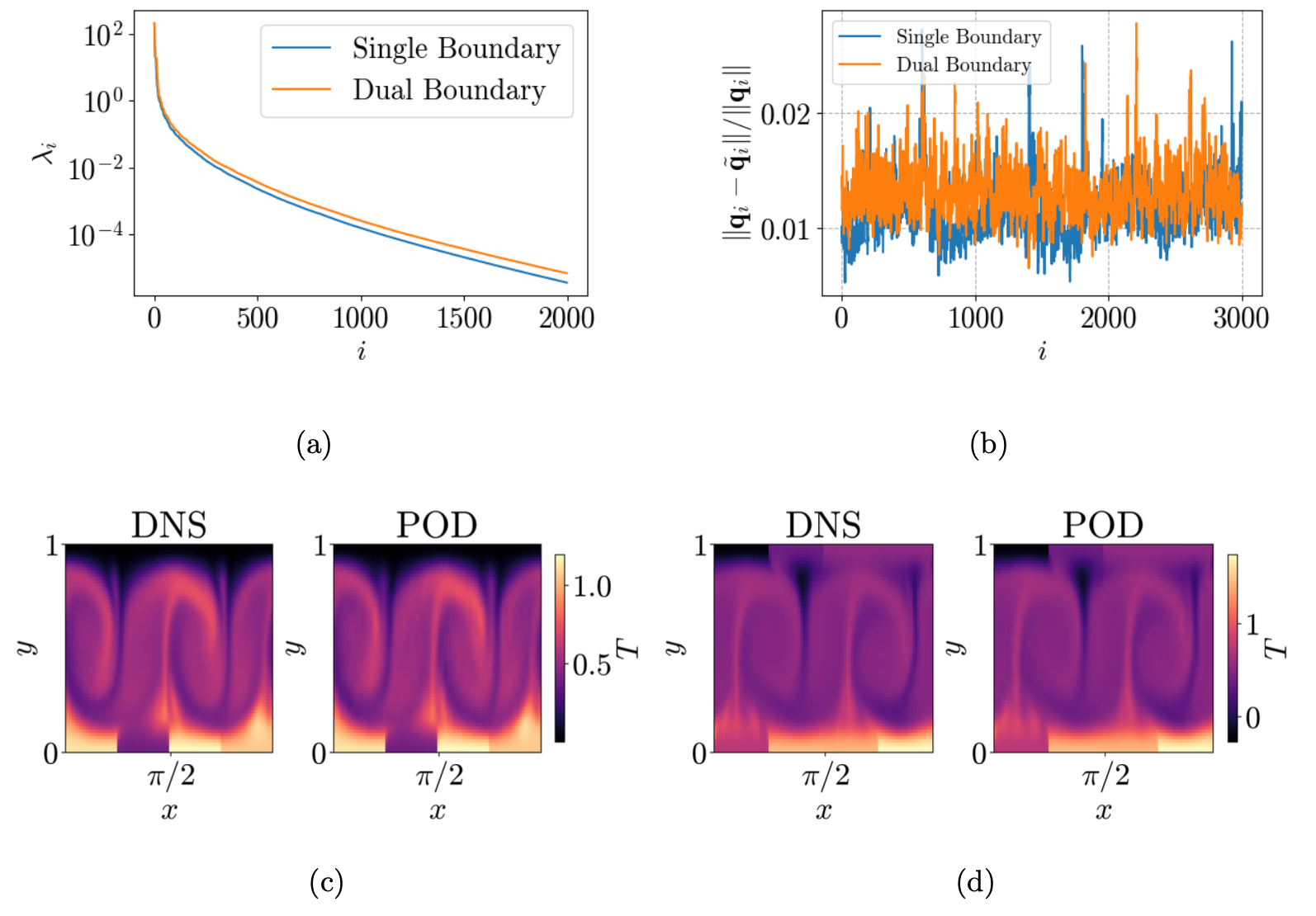}
  \caption{Linear dimension reduction:
  (a) Eigenvalue spectrum.
  (b) Temporal evolution of the reconstruction error \(\| \mathbf{q}_i-\tilde{\mathbf{q}}_i\|_2/\|\mathbf{q}_i\|_2\) on the test set for single and dual actuation cases.
  (c–d) Snapshots of the temperature field  for the  DNS  and POD reconstructions (right) for single- and dual-boundary cases.}
  \label{fig:pod-summary}
\end{figure}

Figure~\ref{fig:pod-summary} provides a summary of the linear dimension reduction for both boundary configurations. Panel~\ref{fig:pod-summary}a shows the eigenvalue spectra for the single- and dual-boundary cases. Both exhibit a steep initial decay, indicating that the flow is dominated by a limited number of energetic modes. The dual-boundary configuration retains slightly more energy in the higher-order modes, reflecting the richer thermal plume interactions and symmetry breaking induced by top-side actuation. 
Figure ~\ref{fig:pod-summary}b presents the temporal evolution of relative reconstruction error on the test set, defined as 
$(\|\mathbf{q}_i - \tilde{\mathbf{q}}_i\|_2)/{\|\mathbf{q}_i\|_2}$,
where $\mathbf{q}_i$ and $\tilde{\mathbf{q}}_i$ denote the original and POD-reconstructed flow snapshots, respectively.  
The error remains at the $\mathcal{O}(10^{-2})$ level for both configurations and is slightly larger for the dual-boundary case, consistent with its higher spectral richness and stronger plume activity.
Figures \ref{fig:pod-summary}c,d compare the reconstructed temperature fields $T$ with  DNS results for representative  single- and dual-boundary control, respectively. In both cases, the leading POD modes successfully reproduce the dominant roll structures and overall temperature distribution. The single-boundary configuration yields smoother, more symmetric rolls, whereas the dual-boundary configuration displays more fragmented and intermittent plume features, consistent with its broader energy spectrum.  

\begin{figure}
    \centering
    \includegraphics[width=\textwidth]{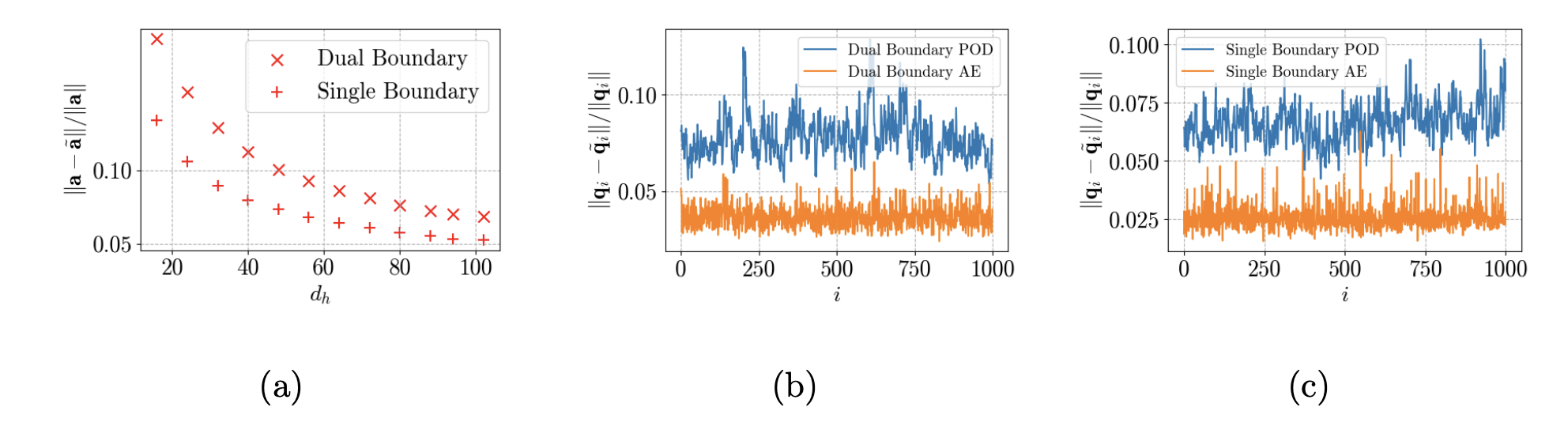}
    \caption{
    Non-linear dimension reduction via autoencoders.
    (a) Average relative error in POD coefficients  as a function of the latent dimension $d_h$.
    (b–c) Temporal evolution of reconstruction errors for the dual- and single-boundary cases at $r=d_h=88$.}
    \label{fig:ae-performance}
\end{figure}

Next, we perform a  nonlinear reduction of the POD coefficients.
Figure~\ref{fig:ae-performance} summarizes the performance of autoencoders for both single- and dual-boundary configurations. Figure~\ref{fig:ae-performance}a shows the variation of the mean relative reconstruction error with the latent dimension $d_h$ (i.e., $    (\|\mathbf{a}_i - \tilde{\mathbf{a}}_i\|_2)/{\|\mathbf{a}_i\|_2}$). For both single- and dual-boundary configurations, the error increases  at small $d_h$, and gradually plateaus beyond $d_h \gtrsim 80$, indicating that additional latent dimensions are needed to capture the nonlinear structure. The initial rapid decay reflects the dominance of a few large-scale coherent features captured by the leading latent variables, while the residual error beyond the plateau is associated with small-scale thermal fluctuations. Based on this convergence, we select $d_h = 88$ for both configurations, which provides a good trade-off between reconstruction accuracy and model dimension.
Figure~\ref{fig:ae-performance}b,c  compare the evolution of relative errors of POD and AE reconstructions at the same effective dimension ($r=d_h=88$). Control experiments in the Appendix show that lower dimensions ($d_h = 16, 40$) result in slower convergence and inferior heat transfer suppression compared to $d_h = 88$. These results indicate that a sufficiently high latent dimension is necessary to capture the critical dynamics required for effective control. The autoencoder consistently achieves lower errors, demonstrating its ability to capture nonlinear correlations among POD modes and thus provide more accurate reduced-order representations under fewer degrees of freedom.

\subsection{Neural ODE for latent dynamics}

\begin{figure}
    \centering
    \includegraphics[width=\textwidth]{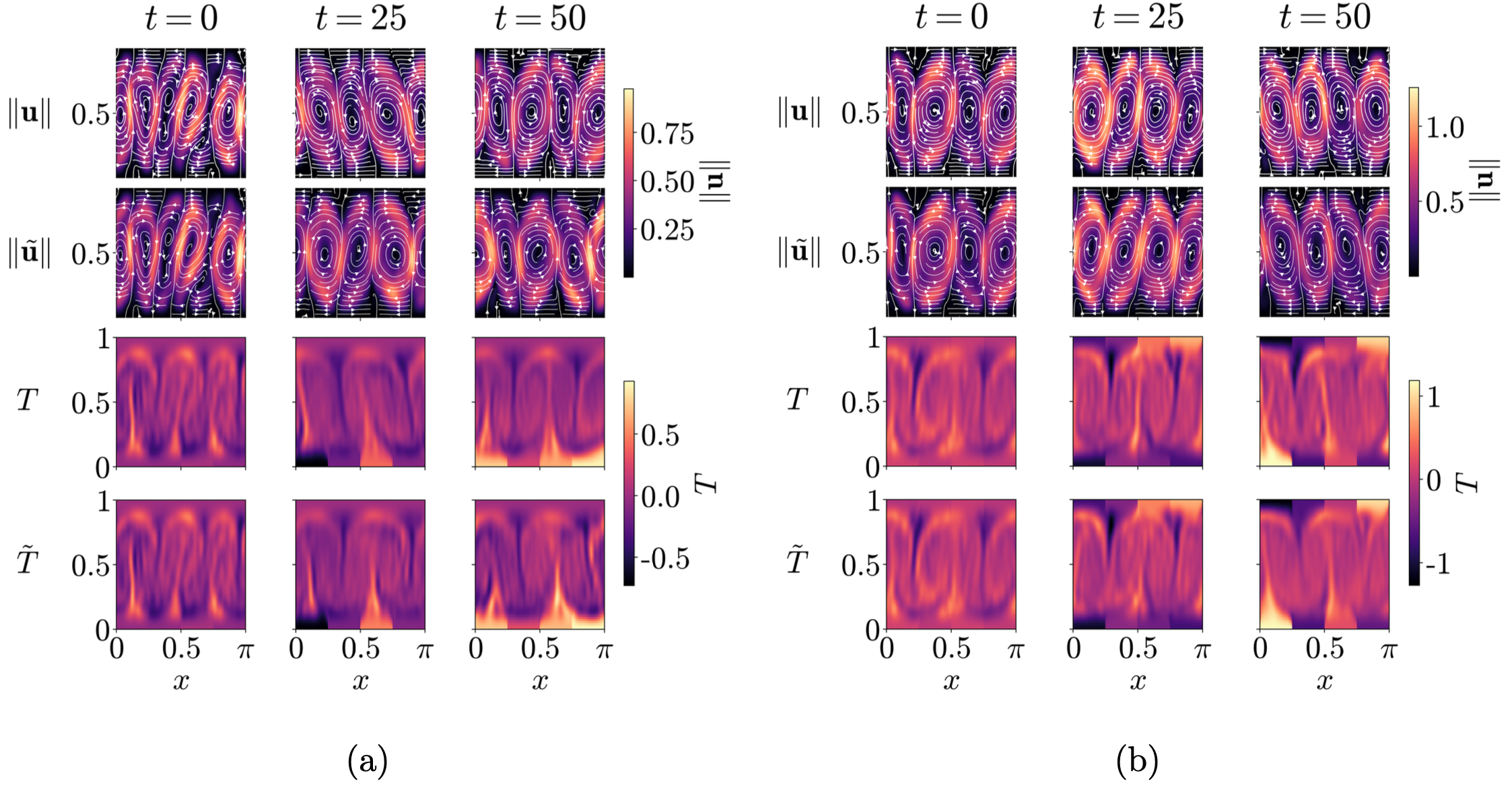}
    \caption{Comparison of DNS and DManD reconstructed fields for (a) single-boundary and (b) dual-boundary control cases. Each panel shows $\|\mathbf{u}\|$ and $T$ at three representative times.}
    \label{fig:node-field-compare}
\end{figure}

Next, we learn an evolution equation in the low-dimensional representation for both cases. Figure ~\ref{fig:node-field-compare}(a) and \ref{fig:node-field-compare}(b) show the magnitude of the velocity and the  temperature fields for the DNS ($||\textbf{u}||$ and $T$) and the 
NODE predictions ($||\tilde{\textbf{u}}||$ and $\tilde{T}$).
Both the single-  and dual-boundary  cases show qualitatively that the DManD model can predict the short time behavior of the system. The models accurately capture  the dynamics, including plume formation, location, and velocity field evolution. This suggests that the selected number of degrees of freedom is sufficient to capture the dynamics in the actuated manifold coordinates.

\begin{figure}
    \centering
    \includegraphics[width=\textwidth]{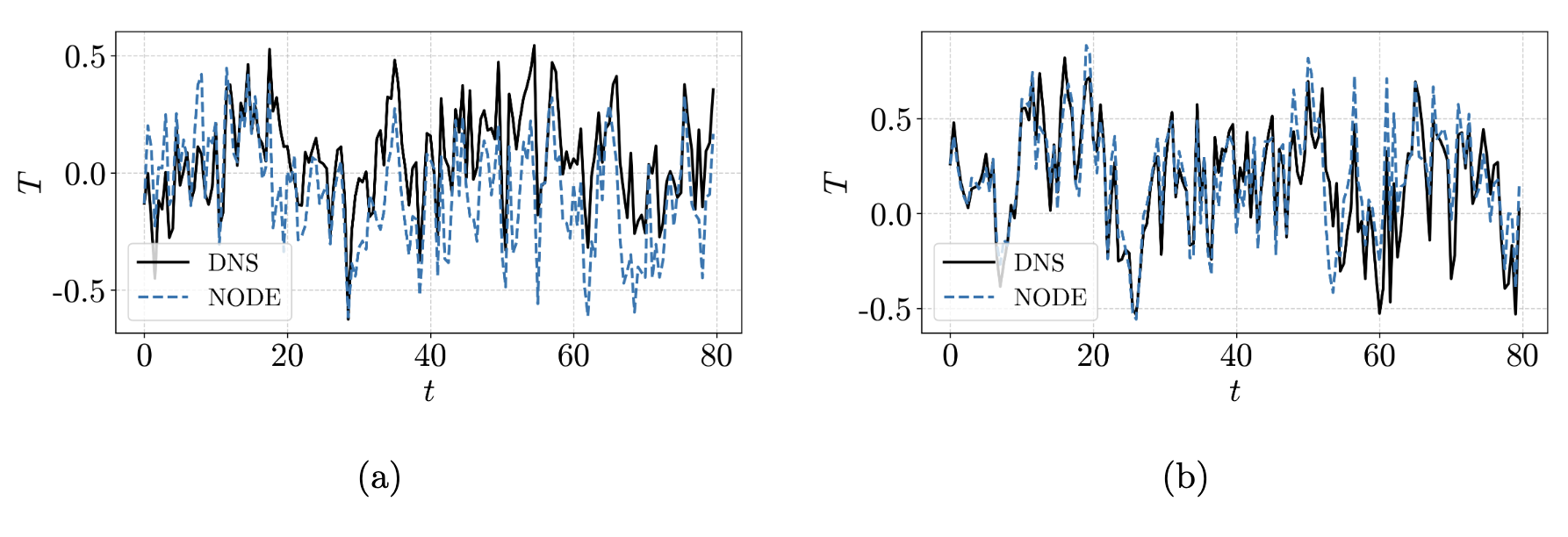}
  \caption{Pointwise time series of temperature $T$ at a probe located at $(x,y)=(\tfrac{\pi}{2},\,0.1)$ for (a) single-boundary and (b) dual-boundary control cases.}
  \label{fig:node-probes}
\end{figure}

To further assess temporal fidelity at dynamically sensitive locations, we monitor a probe at \((x,y)=(\tfrac \pi 2,0.1)\). Figure~\ref{fig:node-probes} shows time traces of \(T\) for both single- and dual-boundary cases. We can see that the NODE tracks the phase and amplitude of the DNS signal well over several control intervals. A slight phase drift develops as the prediction time increases, which is expected for chaotic convection; nevertheless, the qualitative cycle such as the growth and decay of rolls and plume bursts remains synchronized over short windows.

\begin{figure}
    \centering
    \includegraphics[width=\textwidth]{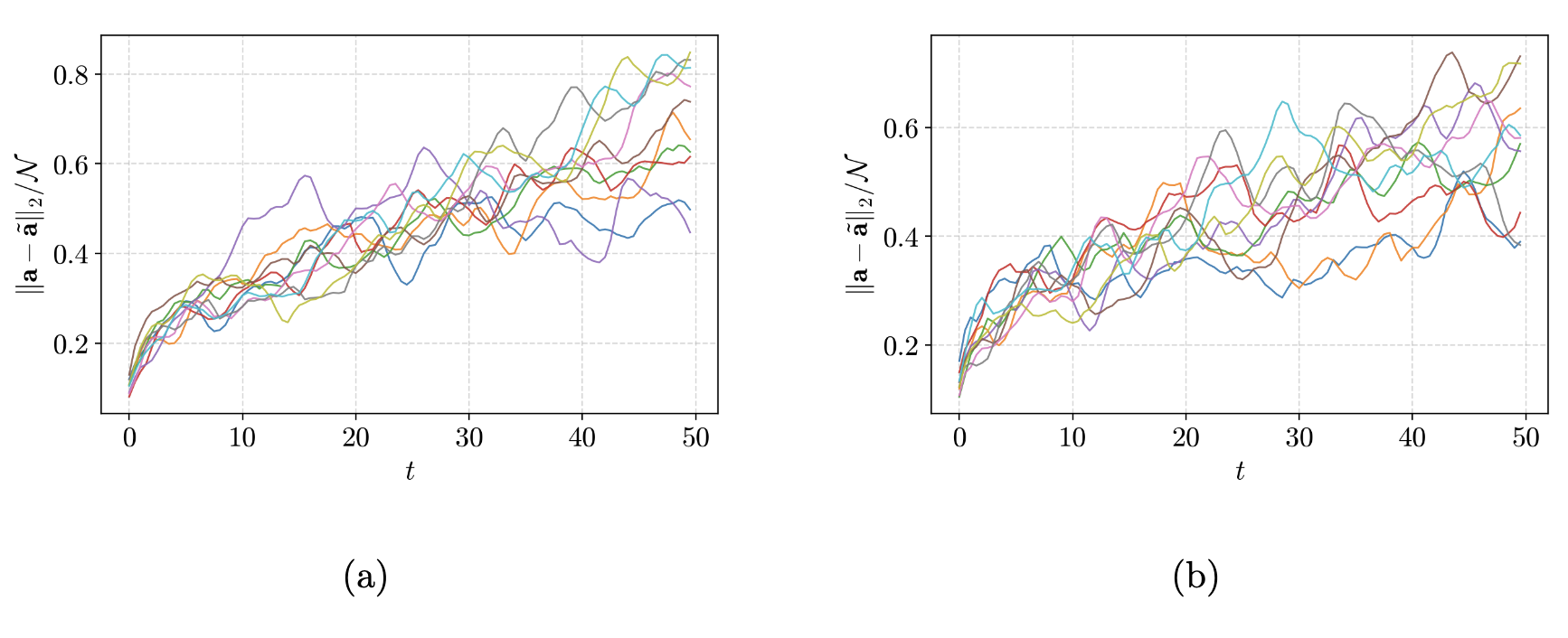}
    \caption{Short-time tracking error of POD coefficients for (a) single-boundary and (b) dual-boundary control cases.}
  \label{fig:node-topk}
\end{figure}

We next quantify  the tracking accuracy directly in the POD coefficient space. Let \(\mathbf{a}(t)\) be the DNS-projected POD vector and \(\tilde{\mathbf{a}}(t)\) the NODE prediction (after AE decoding and POD projection). Following the tracking metric in our short-time analysis, we report the normalized error
\begin{equation}
  \mathcal{E}(t)\;=\;\frac{\big\|\mathbf{a}(t)-\tilde{\mathbf{a}}(t)\big\|_2}{\mathcal{N}}, 
  \qquad 
  \mathcal{N}\;=\;\Big\langle\,\big\|\mathbf{a}(t_i)-\mathbf{a}(t_j)\big\|_2\,\Big\rangle_{\,t_i\neq t_j},
  \label{eq:track_metric}
\end{equation}
where we sets a short-time window to normalize errors by the intrinsic scale of trajectory separation within the attractor. \(\mathcal{E}(t)\) therefore reflects how quickly prediction separates from the ground truth relative to the average of trajectory distances $\mathcal{N}$. Figure~\ref{fig:node-topk} shows the evolution of \(\mathcal{E}(t)\) for ten  randomly chosen  initial conditions. In both cases, the error grows mildly at first, indicating that the predicted and true trajectories remain similar for some time. This demonstrates that the NODE model generalizes well across different initial conditions. 

Finally, we do not report long-time statistics here. Because the system is continually driven by externally updated boundary controls, it does not settle into a stationary regime over our runs; long-horizon averages would mix disparate control-induced regimes and are not physically meaningful for the present controlled setting.

\subsection{DNS evaluation of DManD-RL control strategies}

\begin{figure}
    \centering
    \includegraphics[width=\textwidth]{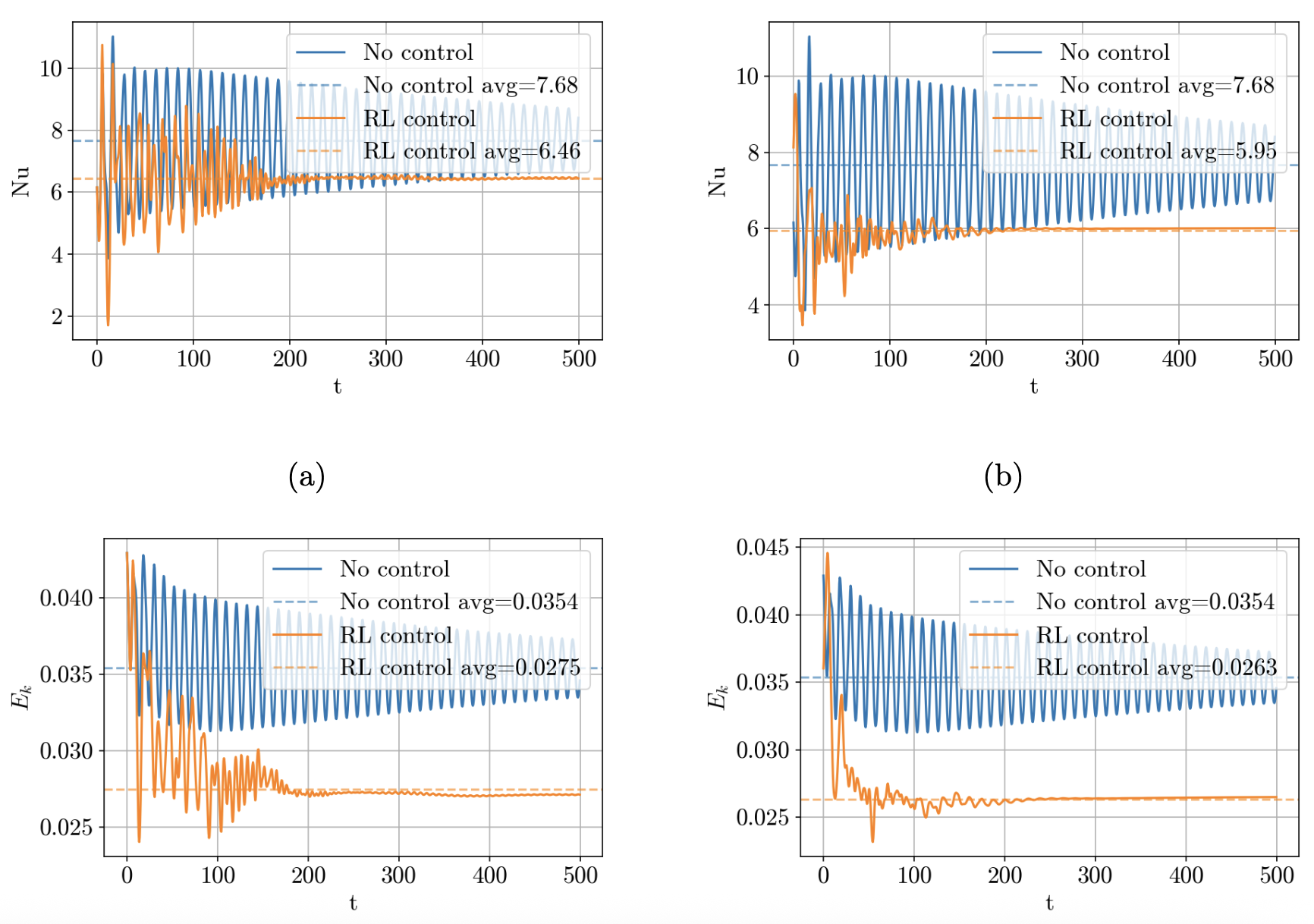}
    \caption{Control policy applied in DNS. Panels (a–b) show the time evolution of the Nusselt number  for the uncontrolled baseline and the RL-controlled cases with (a) single-boundary actuation and (b) dual-boundary actuation. Panels (c–d) show the corresponding  kinetic energy, respectively. 
    }
    \label{fig:nu-comparison-both}
\end{figure}

The performance of the RL control strategies is summarized in Figure~\ref{fig:nu-comparison-both}.
Figures~\ref{fig:nu-comparison-both}a,b compare the instantaneous Nusselt between uncontrolled and RL-controlled simulations. 
In both single- and dual-boundary configurations, the RL controller significantly reduces the amplitude of convective heat transport, suppressing large plume events and smoothing oscillations that decay toward a quasi-steady fixed point, rather than sustaining periodic fluctuations.   
On average, the Nu number decreases from approximately \(7.68\) (uncontrolled) to \(6.46\) and \(5.95\) under single- and dual-boundary control, respectively—corresponding to 15.88\% and 22.53\% reduction in mean heat flux. This reduction is comparable with other state-of-the-art frameworks, which reported reductions around 10\% with a single-agent Proximal Policy Optimization model~\citep{markmann2025controlrayleighbenardconvectioneffectiveness} and up to 22.7\% with a multi-agent model~\cite{vignon2023effective}. While our training relies on full-field data, similar to \citet{markmann2025controlrayleighbenardconvectioneffectiveness}, we recognize that practical applications must rely on sparse sensors. To address this, we  adopt a decoupled strategy in which  a policy is first trained on the latent state $\mathbf{h}$, and a separate network is then trained to estimate $\mathbf{h}$ from wall measurements (similar strategy was done by  \citet{linot2023dynamics}). Since coherent flow structures leave detectable footprints on the boundaries, this method transfers effectively to 2D RBC. As shown in the Appendix, our sensor-based agent significantly reduces $Nu$. Moreover, to assess the robustness of the RL agent against measurement errors, Gaussian noise was added to the DNS fields observed by the agent. The corresponding results are presented in the Appendix. It can be seen that the agent maintains stable and effective control performance even in the presence of Gaussian noise.

To complement the observed reduction in Nu, we assess how the RL controller modifies the flow dynamics by examining the temporal evolution of the kinetic energy defined as 
\[
E_k = \frac{1}{2} \iint |\boldsymbol{u}(x,y,t)|^2 \, dxdy.
\]
The temporal kinetic energy in figure~\ref{fig:nu-comparison-both}c,d further highlights the stabilizing effect of RL on the flow dynamics. Both controllers suppress the large-scale oscillations of \(E_k(t)\), with the dual-boundary controller reaching a lower mean kinetic energy (\(0.0263\)) than the single-boundary case (\(0.0275\)). 
The energy decay trend demonstrates that RL successfully damps the dominant convection mode and maintains a quasi-steady state, consistent with the reduced Nu number. We note that both controllers converge to similar steady states, but the dual-boundary case attains it faster and with reduced transient oscillations, owing to the greater control  provided by actuation at both plates.

\vspace{0.5em}

\vspace{0.5em}

\begin{figure}
    \centering
    \includegraphics[width=\textwidth]{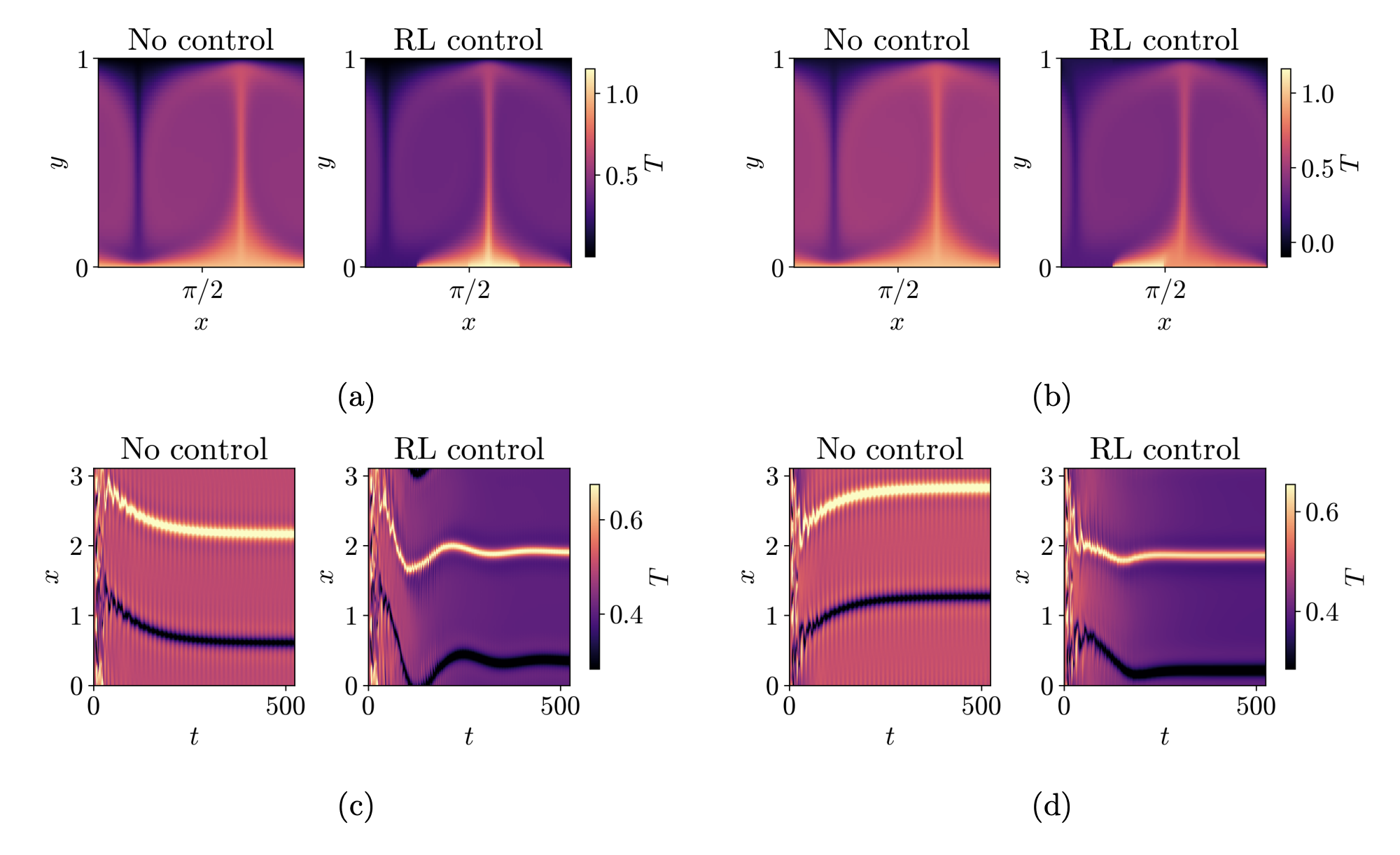}
    \caption{Snapshots of the temperature field at $t = 500$ and corresponding $x\!-\!t$ colormaps at the mid-height plane for one representative initial condition under (a,c) single-boundary and (b,d) dual-boundary control.}
    \label{fig:final-temp-both}
\end{figure}

\begin{figure}
    \centering
    \includegraphics[width=\textwidth]{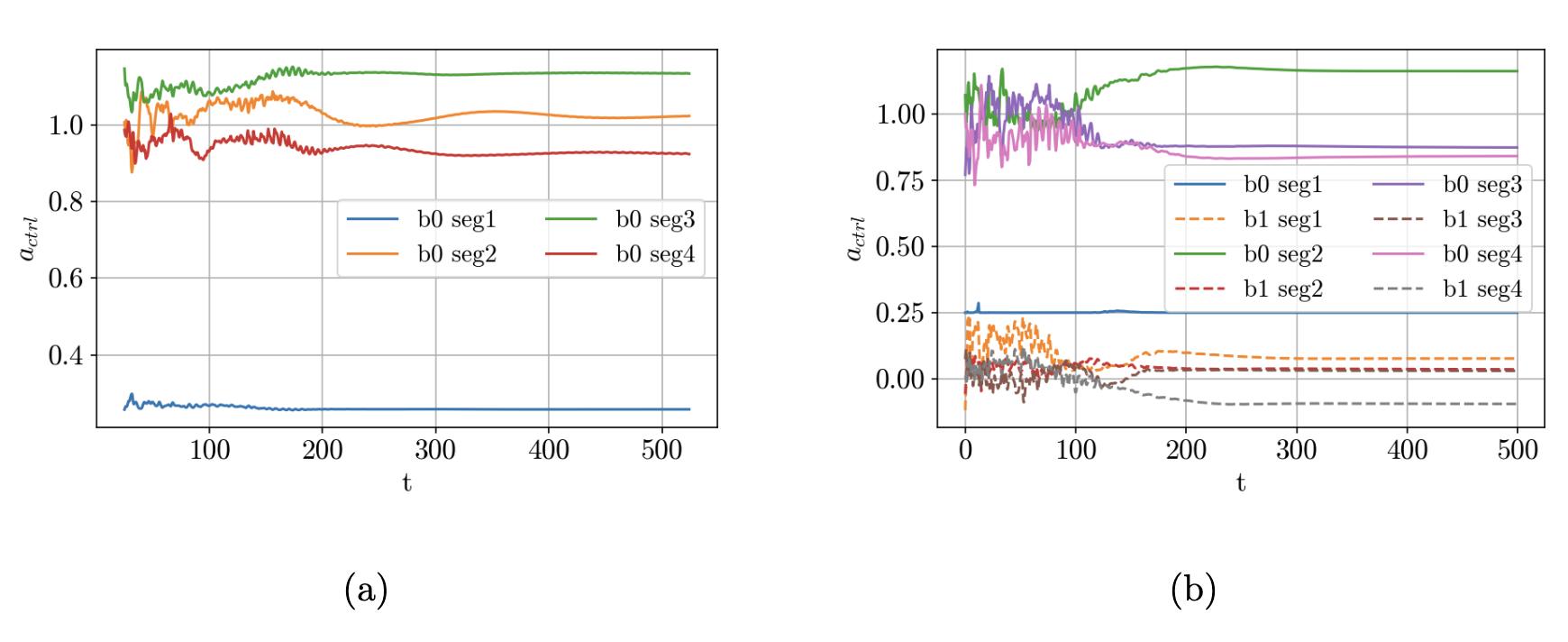}
    \caption{Control signal trajectories \(a_\text{ctrl}(t)\) for (a) single-boundary and (b) dual-boundary control cases. 
    }
    \label{fig:control-evolution-both}
\end{figure}

Figures~\ref{fig:final-temp-both}a,b further illustrate the final stabilized temperature fields  with and without RL control. Relative to the baseline, the controlled cases appear to maintain lower interior temperatures and a more confined hot footprint near the bottom boundary; the primary upwelling near $x\!\approx\!\pi/2$ is narrower and less diffuse.
Figures~\ref{fig:final-temp-both}c–d show the  temporal evolution of the temperature \(T(x,t, z\!=\!0.5)\) as an \(x\!-\!t\) colormap for the same  initial condition under both uncontrolled and RL control. The RL controller suppresses unsteady convection and drives the flow towards a steady state, as evidenced by the absence of temporal thermal variations for $t>300$.  Figure~\ref{fig:control-evolution-both} shows the temporal evolution of the control signals applied to the bottom boundaries (\(b_0\)) and, for the dual case, the top boundaries (\(b_1\)). 
In both configurations, the control segments converge toward nearly constant steady values, consistent with a quasi-static feedback policy that stabilizes the flow through a persistent base heating pattern. 

\begin{figure}
    \centering
    \includegraphics[width=\textwidth]{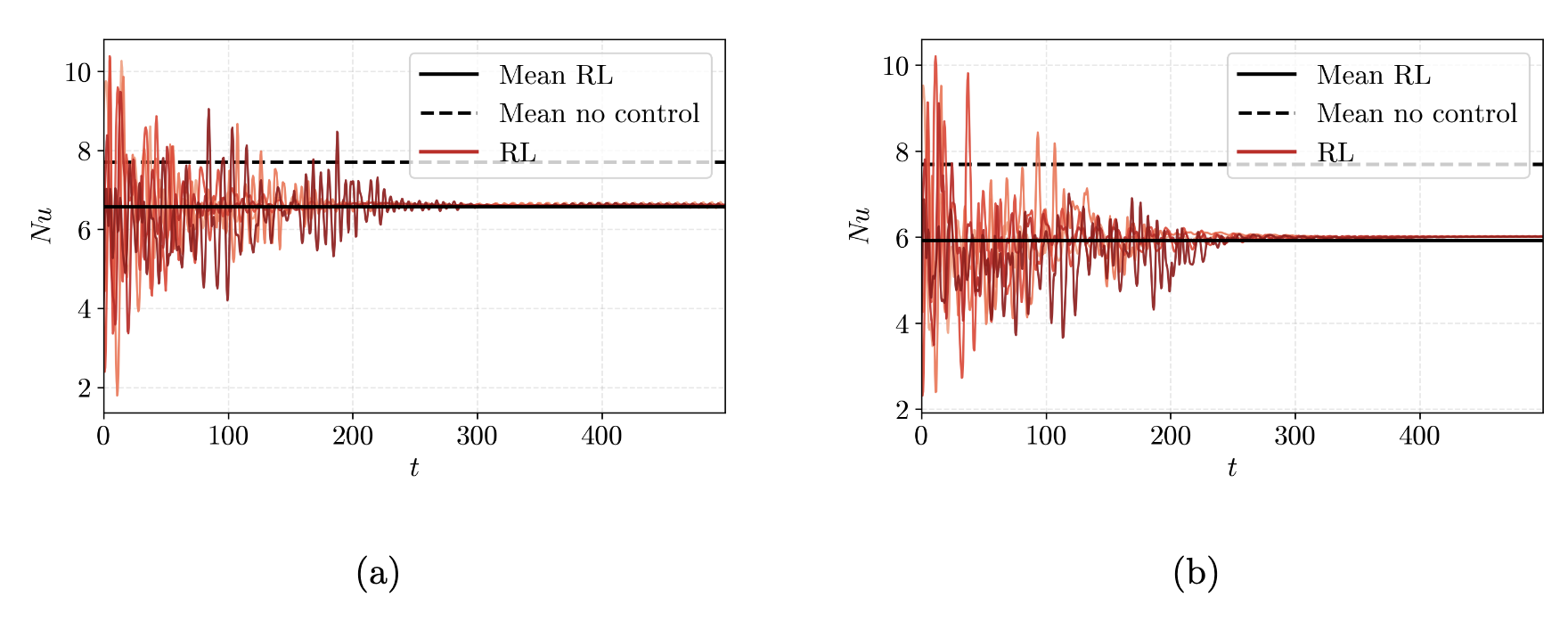}
    \caption{Time evolution of the Nusselt number for ten randomly chosen initial conditions under (a) single-boundary and (b) dual-boundary control.}
    \label{fig:nu-10ics}
\end{figure}

Finally, Figures~\ref{fig:nu-10ics} plot Nu(t) for ten randomly selected initial conditions under single- and dual-boundary RL control, respectively, together with their uncontrolled counterparts.  In all cases, the controller reduces the mean Nusselt number and suppresses temporal variability, driving the system toward a more steady regime. Notably, in all the cases, the controlled flows converge to the same steady state. 

Our simulations showed that the Nu number was reduced by 16–23\% under RL control compared to the uncontrolled baseline. Both single- and dual-boundary strategies achieved substantial stabilization. On an Apple M3 (16 GB), our DManD-RL trains 31.6 times faster than classical DNS-based RL. It is 9.68 ms vs 306 ms per control cycle. Projected to $10^6$ control cycles, DManD-RL requires approximately ~2.7 hours compared with about ~85 hours for DNS-based RL.

\subsection{Physical interpretation of the $Nu$ Reduction}

Given the marked reduction in heat transport and kinetic energy achieved by the DManD–RL controller, this section  examines the underlying physical mechanisms learned by the controller by focusing on the 
near-wall dynamics and large-scale organization.

\begin{figure}
    \centering
    \includegraphics[width=\textwidth]{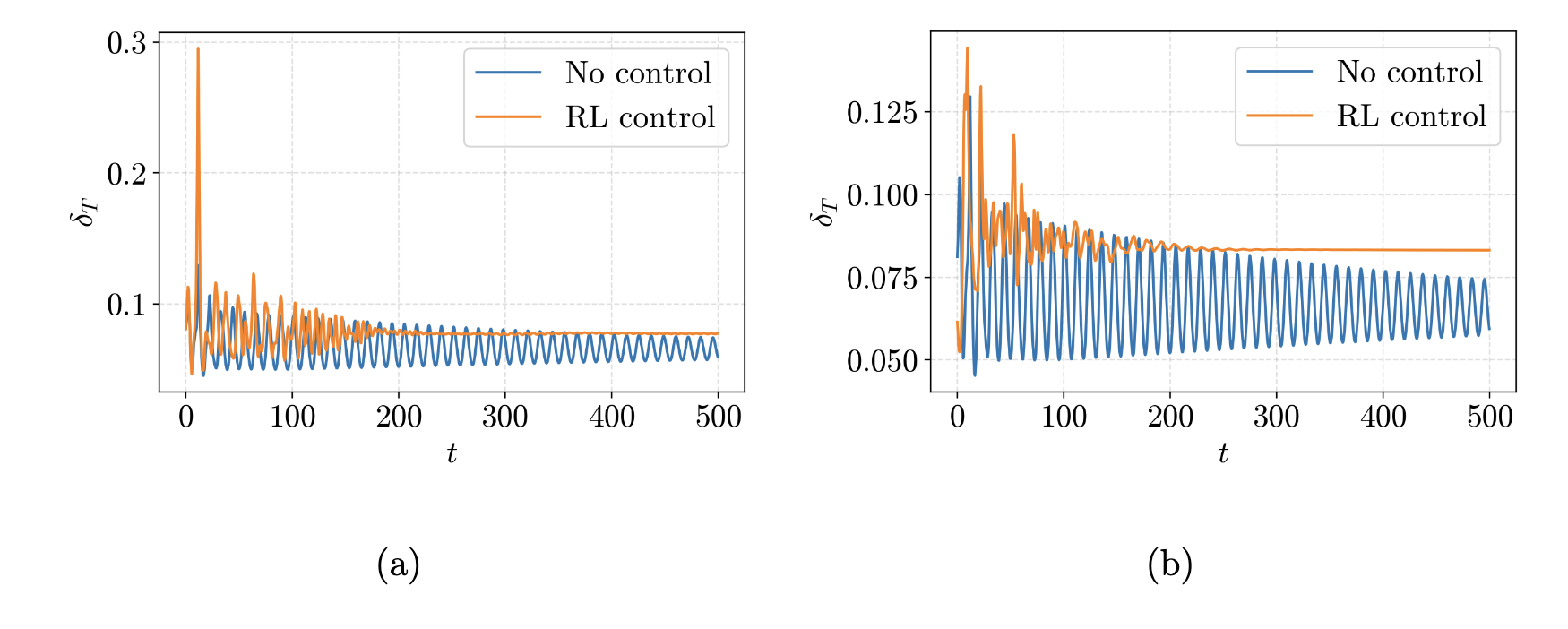}
    \caption{Thermal boundary layer thickness under RL control for (a) single-boundary and (b) dual-boundary control cases.}
    \label{fig:BL-spectrum-both}
\end{figure}

There has been extensive discussion in the literature regarding the connection between thermal boundary-layer instability and plume formation~\citep{malkus1954heat, howard1966convection, puthenveettil2011length}.
As suggested by~\citet{qin2024formation}, the generation of thermal plumes can be interpreted through the evolution of the  thermal boundary layer thickness.
To examine this process, we first analyze the temporal dynamics of the instantaneous thermal boundary layer thickness, defined as 
\[
\delta_T(t) \;=\; H/(2~\mathrm{Nu}(t)|_{y=0})
\] 
following the previous work by ~\citet{ahlers2009heat, shishkina2009mean, chillà2012new}. Figure~\ref{fig:BL-spectrum-both} shows the boundary layer thickness for both actuation configurations. 
After the controller is activated, $\delta_T$ rapidly relaxes to a quasi-steady state characterized by a larger mean thickness, but significantly weaker temporal fluctuations compared with the uncontrolled case. 
The thickening of the boundary layer reflects the reduction in convective heat transport induced by the controller, since a lower heat flux corresponds to a weaker wall temperature gradient.
According to~\citet{qin2024formation}, the evolution of $\delta_T(t)$ in the uncontrolled regime can be divided into four stages: 
(i) boundary-layer growth, (ii) destabilization and plume emission once 	
$\delta$   exceeds a critical threshold, (iii) rapid thinning following plume detachment, and (iv) gradual regrowth. The cyclic repetition of this process directly links $\delta$ fluctuations to plume formation. Under RL control, this oscillatory cycle is strongly attenuated, indicating that the controller stabilizes the boundary layer and suppresses plume activity by maintaining a thicker, more stationary thermal layer.

\begin{figure}
    \centering
    \includegraphics[width=\textwidth]{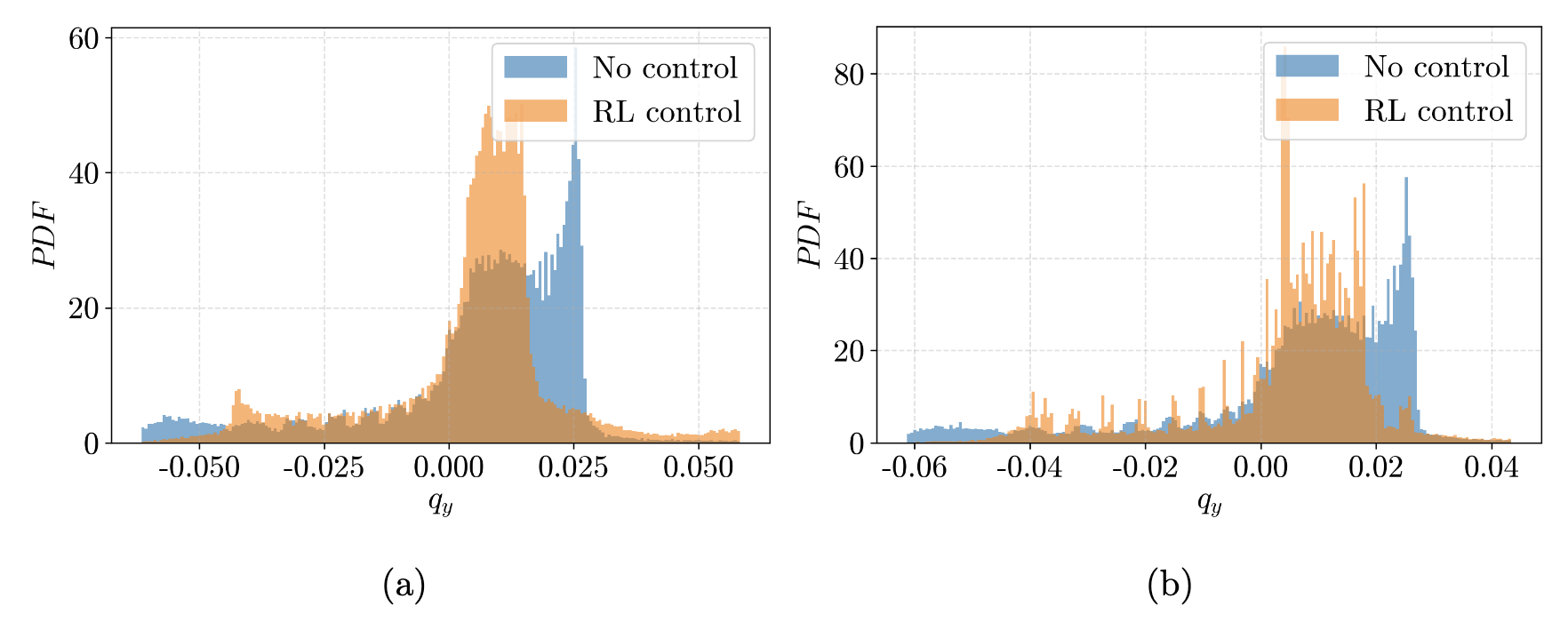}
    \caption{
    PDFs of local instantaneous vertical heat flux \(q_y = u_y T\)
    at \(y = 0.075\) within the thermal boundary layer for (a) single-boundary and (b) dual-boundary control cases.}
    \label{fig:PDF-qy}
\end{figure}

Figure~\ref{fig:PDF-qy} shows the probability density functions (PDFs) of the local instantaneous vertical heat flux  (i.e., $q_y = u_y T$) 
at \(y = 0.075\), which lies within the thermal boundary layer. The PDFs exhibit a pronounced positive skewness, which is considerably stronger in the uncontrolled case than in the RL-controlled flow. As discussed by~\citet{PhysRevE.86.056315}, such positive skewness originates from the asymmetric occurrence of upward-moving warm and downward-moving cold plumes
The reduction of skewness under RL control  indicates that the actuation weakens the plume emission and penetration events within the boundary layer, leading to diminished convective transport and a lower global Nusselt number. Interestingly, while the single-boundary control reduces the overall skewness, it occasionally produces rare positive flux events, reflecting sporadic strong plumes. In contrast, the dual-boundary control does not exhibit such intermittent bursts. This difference may explain why the dual-boundary control achieves a more consistent suppression of plume activity and consequently a greater reduction in the global Nu number.

\begin{figure}
    \centering
    \includegraphics[width=\textwidth]{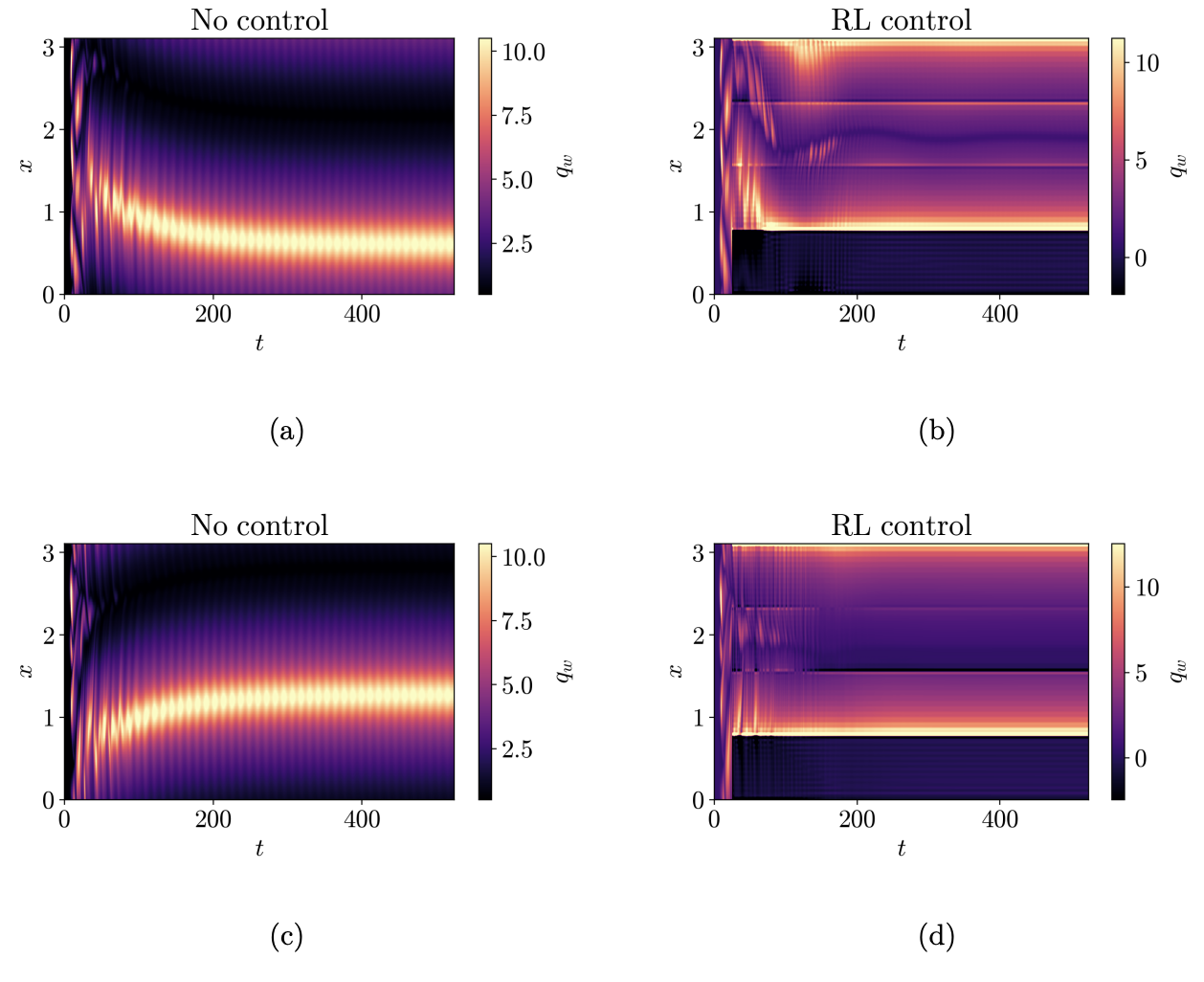}
    \caption{
Space--time evolution of the wall heat flux 
for (a,b) single-boundary and (c,d) dual-boundary control cases.
    }
    \label{fig:qw-spacetime-both}
\end{figure}

Moreover, we examine the space–time evolution of the instantaneous wall heat flux ($q_w(x,t)= -(\partial T / \partial y)\big|_{y=0}$ 
 ) shown in 
Figure~\ref{fig:qw-spacetime-both}. The $x$–$t$ diagrams visualize how the wall heat transfer evolves simultaneously in space and time, providing a signature of the plume impact pattern~\citep{PhysRevLett.90.074501, PhysRevE.86.056315}. In the uncontrolled cases, $q_w(x,t)$ exhibits a bright obliquely drifting streaks, each corresponding to the emergence or impingement of a warm or cold plume on the wall~\citep{PhysRevE.86.056315}.
The persistence and inclination of the streaks reveal an unsteady circulation roll that continuously sweeps hot and cold regions along the wall. Under RL control, the bright bands disappear, replaced by nearly horizontal, piecewise-uniform plateaus of $q_w(x,t)$. This pattern suggests that the wall heat flux becomes spatially segmented and temporally steady, consistent with the suppression of plume impingement and the stabilization of near-wall thermal dynamics. 

\begin{figure}
    \centering
    \includegraphics[width=\textwidth]{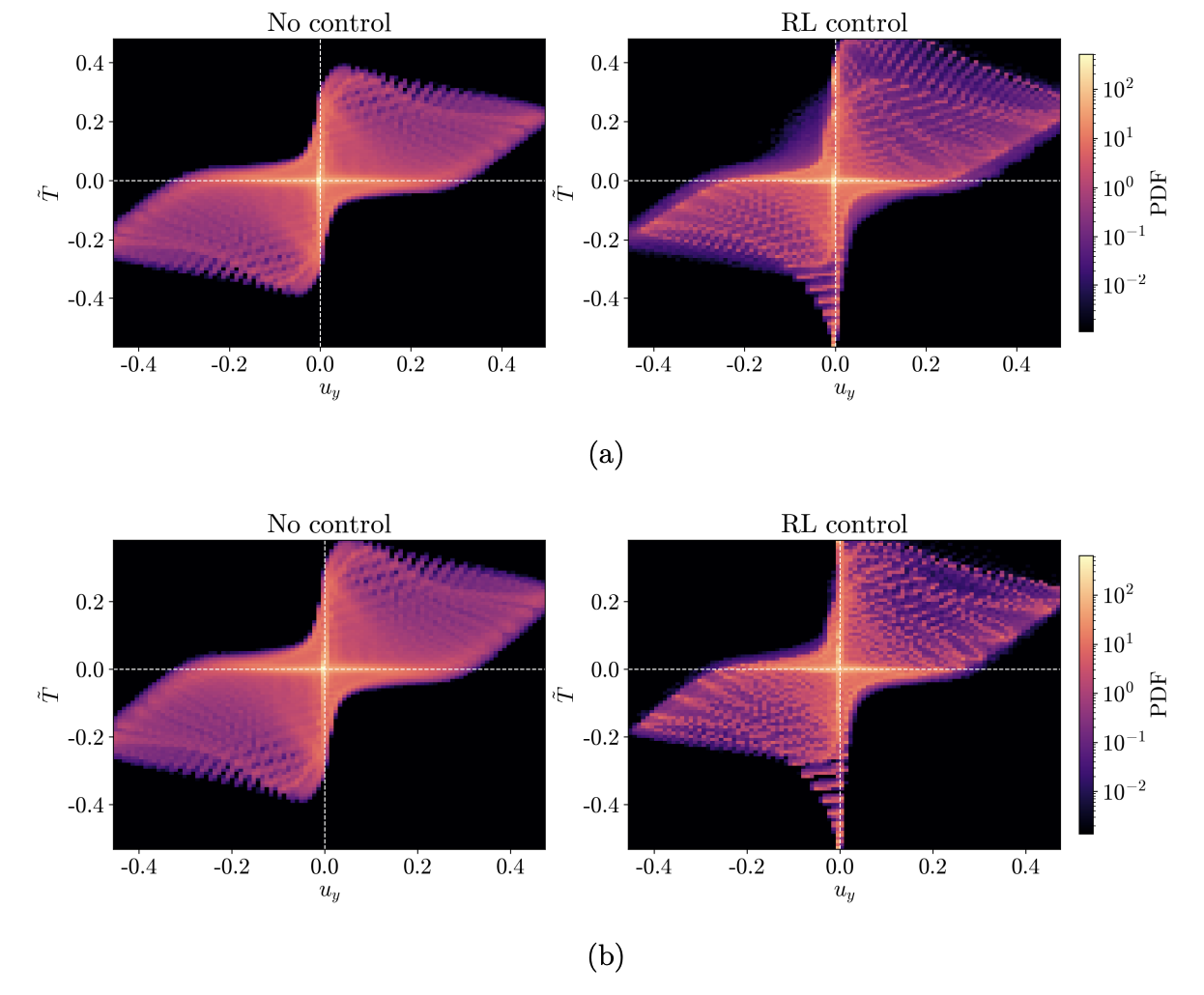}
    \caption{Joint PDFs of the standardized fluctuations $(\tilde{u}_y, \tilde{T})$ for (a) single-boundary and (b) dual-boundary control cases.} 
    \label{fig:quadrant-pdf-all}
\end{figure}

Finally, we quantify the instantaneous coupling between vertical motion and thermal deviation in the full domain using the joint PDF of $(u_y, \tilde{T})$, where $\tilde{T}$ is temperature fluctuation  ~\citep{PhysRevFluids.3.013501, PhysRevFluids.5.113506}. Each point in the plane represents a local convective event sampled over all space and time in the statistically steady regime (typically the last 50\% of the simulation time).
At each snapshot, every grid point $(x_i,y_j)$ contributes one sample pair, so that the empirical PDF is
\begin{equation}
  P(u_y,\tilde{T}) \;=\; \frac{1}{N_x N_y N_t}
  \sum_{i=1}^{N_x}\sum_{j=1}^{N_y}\sum_{n=1}^{N_t}
  \delta\!\big(u_y - u_y(x_i,y_j,t_n)\big)\,
  \delta\!\big(\tilde{T} - \tilde{T}(x_i,y_j,t_n)\big),
  \label{eq:joint-pdf}
\end{equation}
where $\delta(\cdot)$ the Dirac delta fuction.
Positive $u_y$ indicates upward motion, while positive $\tilde{T}$ denotes fluid warmer than the mean temperature. The joint distribution therefore reveals the dominant mechanisms of heat transport. In turbulent RBC, the PDF typically exhibits  pronounced elongation along the first and third quadrants~\citep{PhysRevFluids.3.013501}, where $u_y$ and $\tilde{T}$ share the same sign. These regions correspond to upward transport of hot plumes and downward transport of cold plumes, both contributing positively to the convective heat flux and hence to the instantaneous Nu number.  
Under RL control, the joint PDF becomes elongated along the $\tilde{T}$ axis, indicating more frequent near-zero $u_y$ events. This trend resembles the effect of reducing the aspect ratio $\Gamma = L_x/L_y$~\citep{PhysRevFluids.3.013501}. Geometrical confinement has two competing consequences: stronger viscous drag slows down the vertical velocities and weakens convective heat transport, whereas more coherent plumes that form can locally enhance it~\citep{PhysRevLett.111.104501}. In our simulations, the RL controller divides the domain into four control patches, effectively introducing a piecewise confinement analogous to reducing $\Gamma$. We observe segmented behavior near the boundaries, where large gradients of velocity and temperature produce additional drag effect -- similar to the sidewall friction under geometrical confinement. 
The elongation of the joint PDFs appears only within a narrow band around  $\tilde{T} = 0$, suggesting that the suppression of vertical motion dominates over plume intensification. This  occurs because the segmented control pattern increases the effective viscous drag acting on the walls, thickening the thermal boundary layer and reducing the Nu number, consistent with~\citet{doi:10.1073/pnas.2403699121}. Unlike strongly confined flows, the effect of coherent plumes is not so significant, as this typically occurs when the LSC is broken under strong geometrical confinement~\citep{PhysRevLett.111.104501}, which is not observed in our results (see Figure~\ref{fig:final-temp-both}). \citet{PhysRevLett.111.104501} further reported that Nu decreases as $\Gamma$ is reduced from 0.6 to 0.3, but increases again when $\Gamma < 0.3$ due to enhanced plume coherence. In our configuration, the effective local aspect ratio is $\Gamma_{\mathrm{eff}} = \pi/4 \approx 0.79 > 0.6$,
 corresponding to a moderately confined regime where Nu still decreases with decreasing $\Gamma$. These findings suggest that the RL controller suppresses heat transport through a  confinement-like mechanism. For narrower domains, fewer control segments may be desirable to prevent excessive confinement.


Overall, the DManD–RL controller effectively suppresses plume-driven convection by stabilizing the thermal boundary layer and inducing a confinement-like effect through segmented control. Consequently, the system relaxes toward a quasi-steady state characterized by markedly reduced convective activity.

\section{Conclusion}
\label{sec:conclusion}

In this paper, we efficiently developed a control strategy to reduce the heat transfer in Rayleigh-Bénard convection  via  temperature actuation using  the DManD-RL framework.
By combining POD and autoencoders, we obtained an 88-dimensional manifold representation of the data, whose dynamics is modeled by a NODE. The resulting DManD model qualitatively reproduces the key turbulent features such as plume formation and temporal evolution, with good short-time predictive capabilities. By using DManD model as the control environment, we demonstrated that RL agents can be trained efficiently in a low dimensional model and subsequently deployed in DNS to suppress convective transport and stabilize the flow. The control policy led to a reduction in the  Nu number  by 16–23\% under RL control compared to the uncontrolled baseline for  both single- and dual-boundary strategies.

The DManD–RL controller learns a physically interpretable strategy that stabilizes convection by modifying the near-wall heat flux in a manner analogous to geometric confinement. Through piecewise actuation along the bottom boundary, the policy partitions the wall into quasi-independent segments that locally increase viscous resistance and suppress vertical motions. This segmentation weakens plume ejection and mixing within the thermal boundary layer, leading to thicker, more stable layers and a significant reduction in global heat transport. Rather than amplifying coherent plumes or reorganizing the large-scale circulation, the controller  primarily  damps the wall-driven instabilities that initiate convective bursts. The resulting flow is characterized by reduced temporal fluctuations, spatially steady heat flux patterns, and convergence toward a quasi-steady equilibrium.

DManD-RL has previously been demonstrated to work in systems such as the 1D Kuramoto–Sivashinsky equation and turbulent minimal Couette flow.
Its application here to 2D turbulent Rayleigh–Bénard convection   illustrates the framework’s versatility in handling buoyancy-driven flows governed by different physical mechanisms.
Compared with previous RBC studies with RL such as~\citet{vignon2023effective} and~\citet{markmann2025controlrayleighbenardconvectioneffectiveness}, the present work trains control policies at substantially higher $\mathrm{Ra}$, where buoyancy-driven instabilities are stronger and direct numerical simulation becomes prohibitively expensive.
This demonstrates both the efficiency and the physical interpretability of DManD–RL for controlling turbulent convection in strongly nonlinear flow regimes. Looking ahead, future efforts will focus on applying the method to even higher Rayleigh numbers and implementing more practical boundary-actuation schemes.
Such extensions will further broaden the applicability of DManD–RL to complex turbulent flows and other high-dimensional dynamical systems.\\

\noindent{\bf Data Availability.}
The data and codes used in this study are openly available at: \url{https://github.com/CFTL-Illinois/DManD-RL-for-RBC}.\\

\noindent{\bf Acknowledgments}. This research used the Delta advanced computing and data resource which is supported by the National Science Foundation (award OAC 2005572) and the State of Illinois.

\section*{Appendix}
\subsection{Sensitivity of control performance to latent dimension }

\begin{figure}
    \centering
    \includegraphics[width=\textwidth]{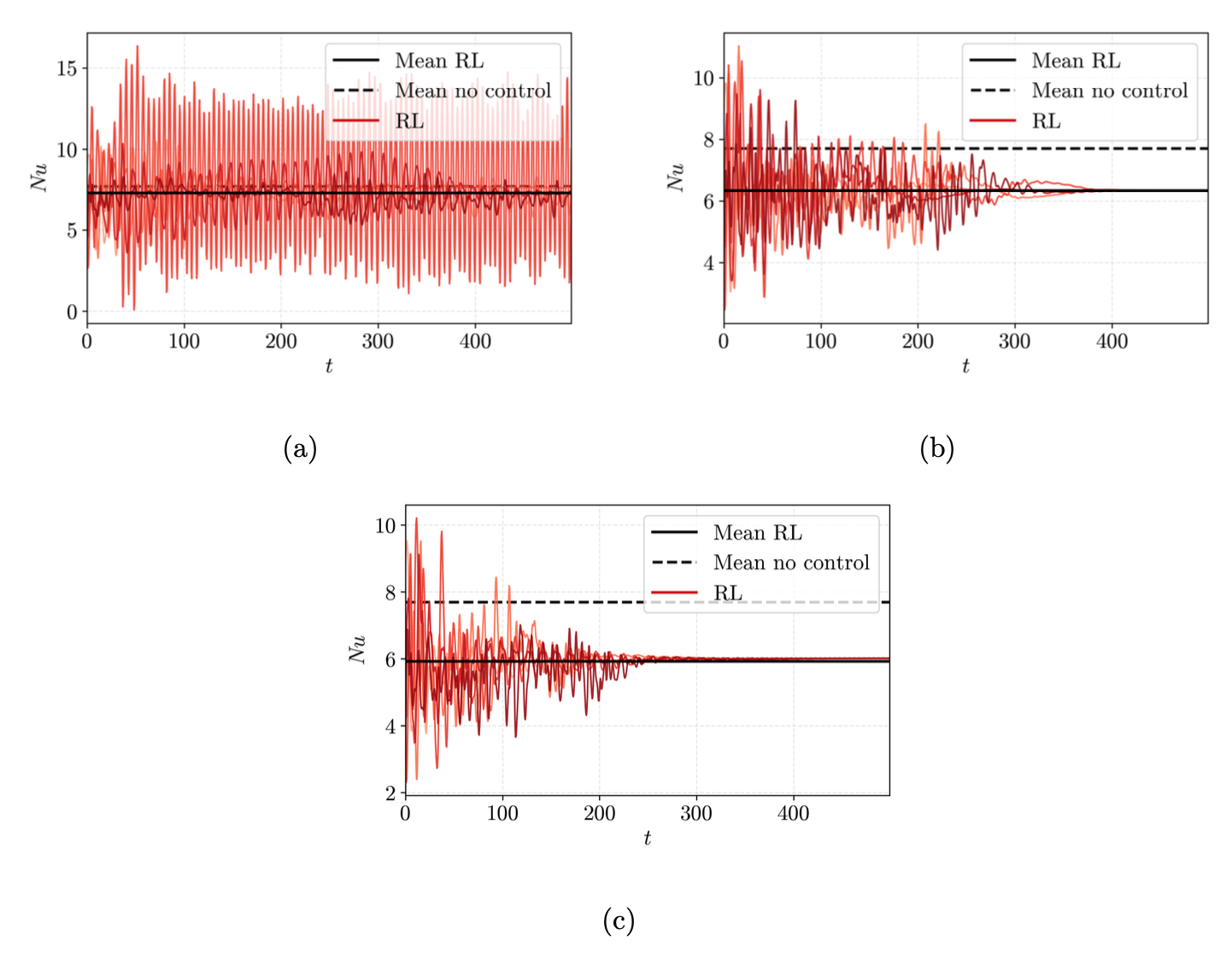}
    \caption{Time evolution of the Nusselt number for ten randomly  initial conditions under dual-boundary control with (a) $d_h = 16$ (b) $d_h = 40$ and (c) $d_h = 88$.}
    \label{fig:appd1}
\end{figure}

While accurate state reconstruction is not a prerequisite for effective reduced-order control (as shown by \citet{Ilak}), the latent representation must retain the dynamically relevant degrees of freedom. In the context of reinforcement-learning-based control, this requirement is particularly important for achieving robust performance across different initial conditions. To assess whether a reduced latent dimension can still support effective control, we examine the sensitivity of the control performance to the size of the latent space. This analysis is included to justify the architectural choice adopted in the main text and to identify the minimum latent dimension required for reliable control.
 
Figure~\ref{fig:appd1} presents the time evolution of the $Nu$ for ten randomly selected initial conditions under dual-boundary control with $d_h = 16$, $40$, and $88$. For $d_h = 16$, the RL agent fails to suppress heat transfer effectively, with $Nu$ trajectories exhibiting large fluctuations. At $d_h = 40$, although the control performance improves, the convergence is notably slower, and the stabilized $Nu$ remains at a higher level compared to the $d_h=88$ case. In contrast, $d_h = 88$ achieves the fastest convergence and the lowest stable $Nu$ across all initial conditions. These results demonstrate that reducing $d_h$ below a certain threshold leads to significant performance degradation, confirming that $d_h = 88$ is necessary for optimal control.

\subsection{Control with wall-based sensing}

\begin{figure}
    \centering
    \includegraphics[width=\textwidth]{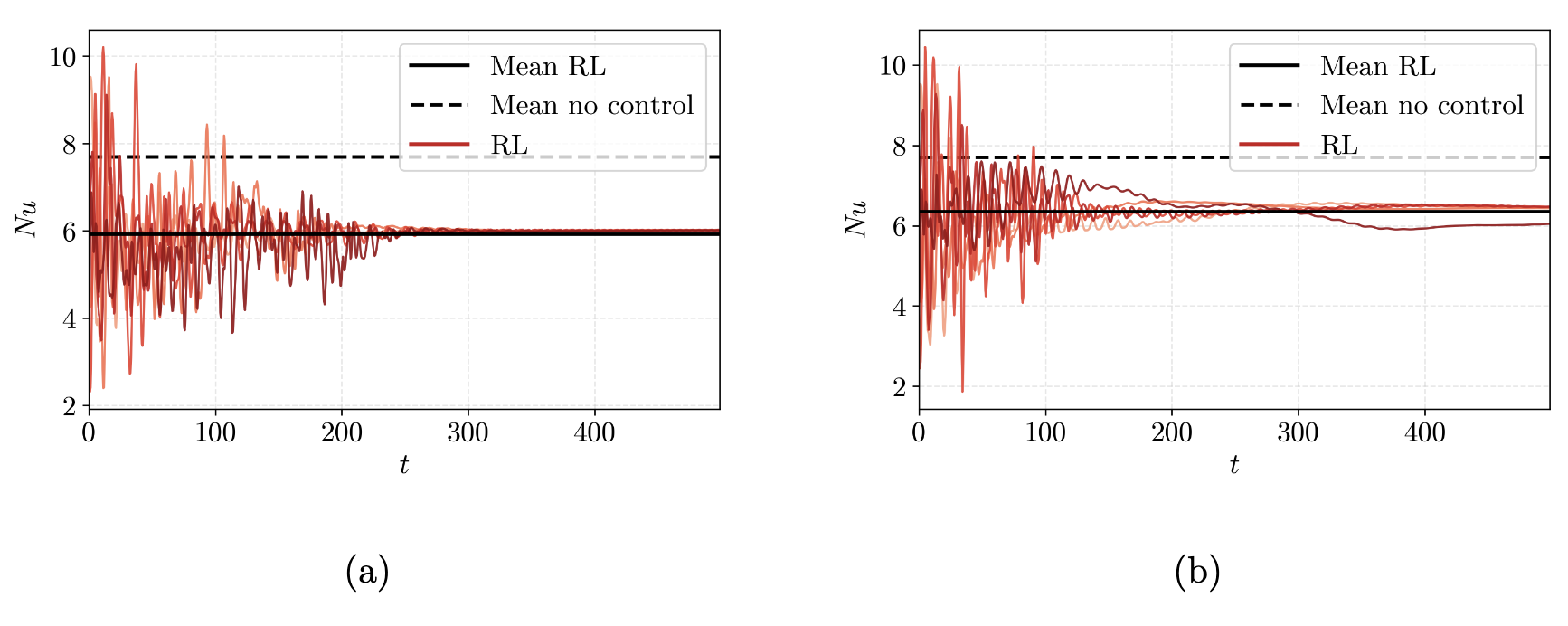}
    \caption{Time evolution of the Nusselt number for ten randomly chosen initial conditions under dual-boundary control with agents trained by (a) latent vector $\boldsymbol{h}$ (b) wall-observations.}
    \label{fig:appd2}
\end{figure}

The results presented in the main text assume access to the full flow field when constructing the latent state used for control. While this assumption is useful for establishing an upper bound on achievable performance, it is not representative of most experimental or practical settings, where measurements are typically restricted to sparse sensors, often located at the boundaries. To address this limitation and assess the practical viability of the proposed framework, we evaluate the control performance when the agent relies exclusively on wall-based observations. This analysis is included to examine the robustness of the control strategy under realistic sensing constraints.

Figure~\ref{fig:appd2} compares the control performance under two different deployment scenarios. Panel~(a) presents the baseline case, in which the full-field latent state $\boldsymbol{h}$ is available to the agent, leading to rapid stabilization and near-optimal control performance. Panel~(b) demonstrates the robustness of the proposed framework when only wall-based observations are provided. In this case, the agent is still able to consistently suppress the mean Nusselt number to the target level across all initial conditions, confirming its practical effectiveness. However, compared with the full-field case, the stabilized $Nu$ remains at a slightly higher level. This degradation is likely attributable to the spatio-temporal delay inherent in boundary-based sensing, whereby information about internal plume dynamics must propagate to the walls before being observed. In addition, the intrinsic reconstruction error of the mapping network introduces further inaccuracies in the estimated latent state.

\subsection{Robustness to measurement noise}

\begin{figure}
    \centering
    \includegraphics[width=\textwidth]{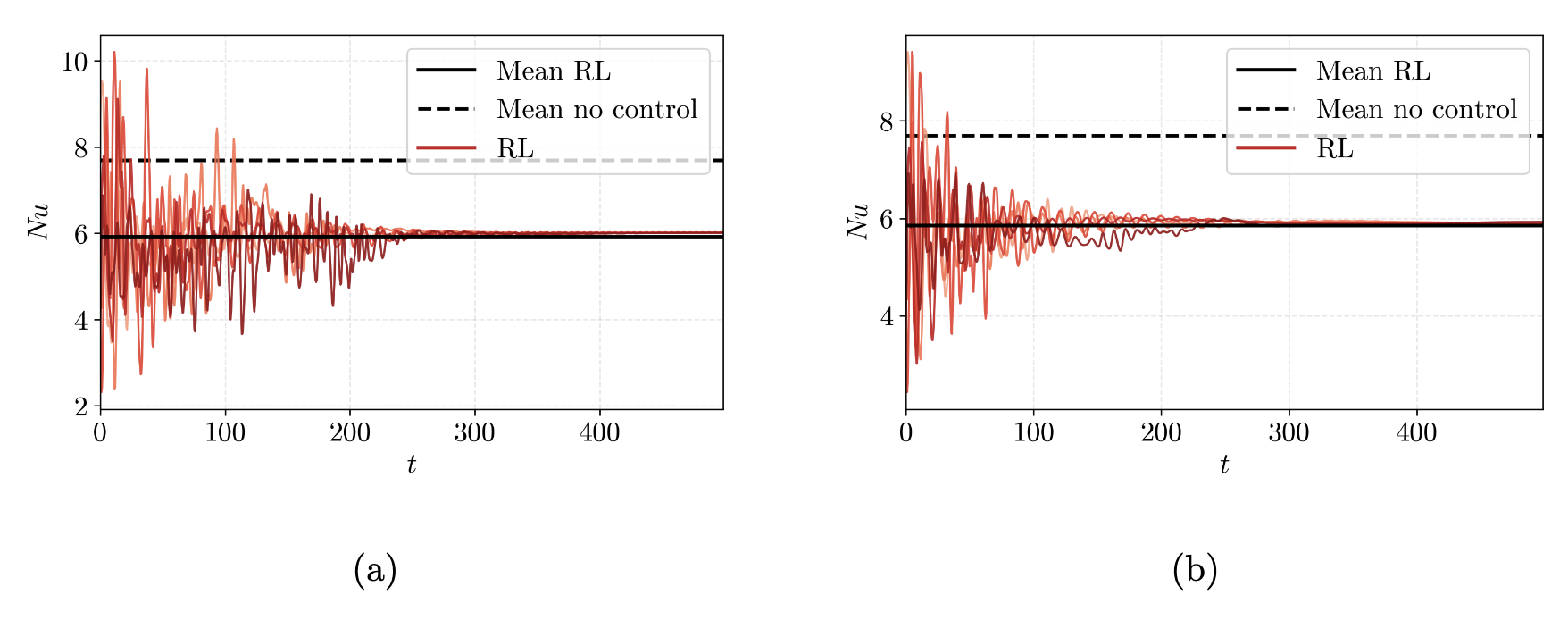}
    \caption{Time evolution of the Nusselt number for ten randomly chosen initial conditions (a) without  noise, and (b) with Gaussian noise of 1\%.}
    \label{fig:appd3}
\end{figure}

To assess robustness to measurement uncertainty, we introduce relative Gaussian noise with an amplitude of $1\%$ into the instantaneous DNS fields observed by the RL agent, emulating measurement errors that may arise in practical experimental settings. Figure~\ref{fig:appd3} shows the time evolution of the Nusselt number, $Nu$, for ten randomly selected initial conditions. Despite the presence of observation noise, the controller consistently drives the system toward a low-$Nu$ state, demonstrating that the learned policy remains effective under noisy sensing.

While a systematic parametric sweep over noise amplitudes is beyond the scope of the present study, the results reported here demonstrate robustness at noise levels comparable to those typically encountered in experimental  measurements. These findings indicate that the proposed  control framework is acceptable to moderate observational uncertainty and does not rely on noise-free state information to achieve effective control.

\bibliography{sample}

\end{document}